\documentclass[prd,aps,twocolumn,a4paper,floatfix,nofootinbib]{revtex4}

\usepackage{lipsum}
\usepackage{graphicx,psfrag}
\usepackage{enumerate}
\usepackage{mathrsfs}
\usepackage{amsmath,amsfonts,amssymb}
\usepackage{enumerate}
\usepackage{hyperref}
\usepackage{float}
\usepackage{ulem}

\newcommand{\p}{\partial}
\newcommand{\Lie}{\mathcal{L}}

\begin{document}

\title{Analytical and numerical treatment of perturbed black holes in
  horizon-penetrating coordinates}

\author{Maitraya K \surname{Bhattacharyya}${}^{1,2}$}
\author{David \surname{Hilditch}${}^{3}$}
\author{K Rajesh \surname{Nayak}${}^{1,2}$}
\author{Hannes R \surname{R{\"u}ter}${}^{4,5}$}
\author{Bernd \surname{Br{\"u}gmann}${}^{4}$}
  
\affiliation{${}^{1}$Indian Institute of Science Education and
  Research Kolkata, Mohanpur 741246, India \\ ${}^{2}$Center of
  Excellence in Space Sciences India, Mohanpur 741246, India\\
  ${}^{3}$Centro de Astrof\'{\i}sica e Gravita\c c\~ao - CENTRA,
  Departamento de F\'{\i}sica, Instituto Superior T\'ecnico - IST,
  Universidade de Lisboa - UL, Av. Rovisco Pais 1, 1049-001 Lisboa,
  Portugal \\ ${}^{4}$Theoretical Physics Institute, University of
  Jena, 07743 Jena, Germany \\ ${}^{5}$Max Planck Institute for
  Gravitational Physics (Albert Einstein Institute), 14476
  Potsdam-Golm, Germany } \date{\today}

\begin{abstract}
The deviations of non-linear perturbations of black holes from the
linear case are important in the context of ringdown signals with
large signal-to-noise ratio. To facilitate a comparison between the
two we derive several results of linear perturbation theory in
coordinates which may be adopted in numerical work. Specifically, our
results are derived in Kerr-Schild coordinates adjusted by a general
height function. In the first part of the paper we address the
questions: for an initial configuration of a massless scalar field,
what is the amplitude of the excited quasinormal mode (QNM) for any
observer outside outside the event horizon, and furthermore what is
the resulting tail contribution? This is done by constructing the full
Green's function for the problem with exact solutions of the confluent
Heun equation satisfying appropriate boundary conditions. In the
second part of the paper, we detail new developments to our
pseudospectral numerical relativity code~\verb|bamps| to handle scalar
fields. In the linear regime we employ precisely the Kerr-Schild
coordinates treated by our previous analysis. In particular, we evolve
pure QNM type initial data along with several other types of initial
data and report on the presence of overtone modes in the signal.
\end{abstract}

\maketitle

\section{Introduction}\label{Section:Introduction}

Black hole perturbation theory~\cite{RW, Zerilli, Vishu, BertiReview,
  Nollert, Kokkotas1999, Konoplya} is an important tool to study
fundamental problems in black hole physics and astrophysics. With the
advent of gravitational wave detectors and direct detection of
gravitational waves from compact
binaries~\cite{LIGO1,LIGO2,LIGO3,LIGO4,LIGO5,LIGO6}, results from
perturbation theory have become increasingly useful in modeling
waveforms for these compact objects. More specifically, the
post-merger part of a binary black hole waveform is dominated by a
linear combination of damped sinusoids with frequencies characteristic
of the mass and spin of the final black hole after merger. These
parameters, called quasinormal mode (QNM) frequencies, have been used
for testing general relativity (GR)~\cite{LIGO7,LIGO8, Li1, Li2, Li3}
and other alternative theories of gravity. It is convenient to use
frequencies derived from the linear theory for these tests. With the
increase in detector sensitivity and the advent of space based
detectors, black hole spectroscopy is poised to become a vital tool
for probing possible deviations from GR in the non-linear regime and
testing the validity of the no-hair conjecture~\cite{Israel,
  Israel2}. The use of overtone modes have been successful in testing
the no-hair conjecture with present gravitational wave
detectors~\cite{Isi}. Stronger tests could be performed with increase
in detector sensitivity and the advent of space based detectors like
LISA.

Several studies have extended the results of the linear theory to
higher orders~\cite{Pullin1, Pullin2, Pullin3, Pullin4, Pullin5} and
to full numerical relativity (NR)~\cite{nonlinear1, nonlinear2,
  nonlinear3, nonlinear4}. However, to the best of our knowledge, a
comprehensive study connecting the results from the linear and the
non-linear theory in the presence of `large' perturbations is still
absent. Several factors, such as second-order QNMs~\cite{Hiroyuki} and
the dependence of tail decay rates on the number of
dimensions~\cite{Bizon2009}, suggest that a full non-linear study may
reveal new physics. It is towards fulfilling this gap that we have
recently been further developing our pseudospectral numerical
relativity (NR) code~\verb|bamps|~\cite{bamps1, bamps2, bamps3,
  bamps4, bamps5, bamps6, bamps7}, our aim being to build a complete
numerical laboratory for perturbation theory experiments. This
challenge requires developing new numerical techniques, notably a more
robust way to handle black hole excision, the ability to extract waves
at null infinity, and data analysis tools to compare linear and
non-linear data.

Motivated by observational and theoretical considerations, we would
thus like a systematic, quantitative comparison between linear,
higher-order perturbative and fully nonlinear solutions with all of
these elements computed in the {\it most} compatible manner. To
facilitate the desired comparison a natural first wish would be to
ascertain the direct, dynamical QNM, and backscattering contributions
to the signal emanating from sufficiently small, but otherwise generic
initial data in both the linear and non-linear contexts. While
searching for an answer to this problem, we found that with few
important notable exceptions, such
as~\cite{ansorgrodrigo,Campanelli_2001, PhysRevD.64.084016}, most
existing calculations have been performed in either Regge-Wheeler or
Schwarzschild coordinates~\cite{LeaverPRD, Andersson1995,
  Andersson1997, Berti2006, CaltechGF, Zhang,
  PhysRevD.84.104002,PhysRevD.38.1040, Frolov:1998wf} and therefore
are not ideal for comparison with full NR results, where simulations
are generally performed in horizon-penetrating coordinates, which may
also be hyperboloidal in nature.

In this paper we therefore restrict our attention to the linear regime
and construct a Green's function in horizon-penetrating Kerr-Schild
coordinates, alternatively named in this context Eddington-Finkelstein
coordinates, offset by an arbitrary height function. The latter can be
used to render the slices hyperboloidal, which will be important for
future numerics. Plain Kerr-Schild coordinates are already needed for
comparison with \verb|bamps|. A simplification of the massless
Klein-Gordon equation to the confluent Heun equation
(CHE)~\cite{Fiziev1, Fiziev2, Fiziev3} is hence provided at the
beginning of section~\ref{Section:Analytical}. We then give a brief
overview of the CHE, its exact solutions and asymptotic solutions at
large radii in section~\ref{Section:HeunSolutions}. A description of
QNM boundary conditions in several coordinate systems along with an
overview of the analytic continuation method used to construct them is
provided in section~\ref{subsection:QNMs}. The exact Green's function
for the problem is constructed using these solutions in
section~\ref{subsection:Exact_Green}. This is then used to compute the
quasinormal mode excitation factors (QNEFs) in
section~\ref{subsection:QNEF}. Separate approximations for the tail at
low, medium and high frequencies are discussed in
section~\ref{subsection:Tails}. The contribution from the direct part
of the signal is discussed in section~\ref{subsection:directGF}.
	
It has been recently suggested that overtone modes may play an
important role in modeling the QNM part of the
signal~\cite{London,Isi,Giesler}. Our analytic calculations and our
Green's function results are valid for arbitrary initial data in an
arbitrary time coordinate related to the Kerr-Schild time by a height
function $h(r)$. In the second part of the paper, for simplicity we
work in the special case of spherical symmetry and evolve various
configurations of a massless scalar field on the Schwarzschild
background in Kerr-Schild coordinates. This presupposed spherical
symmetry ensures that~$l \geq 1$ modes are not excited and QNM ringing
is comprised of the principal frequency and overtone frequencies of
the~$l = 0$ mode. After a description of the numerical setup, the
\verb|scalarfield| project and the initial data within \verb|bamps| in
section~\ref{subsection:Numerical_Setup}
and~\ref{subsection:Numerical_ID}, we perform tests of the tail
results and determine the number of terms that are needed in a data
analysis model to accurately model the numerical results in
section~\ref{subsection:tailnumerics}. One goal of the work is to
investigate the effect of specialized initial data. This is pursued in
section~\ref{subsub:qnmnumerics}, where we employ a method to evolve a
pure QNM solution to obtain an arbitrarily long ringing time near the
horizon. This allows us to evolve and detect overtones or a linear
superposition of them. We then discuss the possibility of detecting
overtone modes from generic initial data and discuss the restricted
circumstances under which this is possible. Finally, we devise a
strategy to prepare specialized initial data with sine-Gaussians which
can be used to obtain long ring-down signals, which improves our
ability to detect the first overtone substantially, at least for
observers far from the horizon. We also demonstrate that irrespective
of the initial data the effect of the branch cut present in the
Green's function construction becomes important during intermediate
and late time ringing. We then present a brief comparison between the
results of our approximate Green's function (for the direct part of
the signal) and the numerics in
section~\ref{subsection:directGF}. Finally in
section~\ref{Section:Conclusions}, we propose a model for QNM ringing
which also incorporates the effect of backscattering, and conclude,
discussing the shortcomings of the present approach.

\section{The wave equation in horizon-penetrating coordinates}
\label{Section:Analytical}

The realistic problem of interest is to evolve an arbitrary
configuration of a massless scalar field in the Schwarzschild
spacetime and study the response of the black hole to it. For
sufficiently weak matter content, we can perform our simulations in
the Cowling approximation, in which the back-reaction from the scalar
field on the metric is considered negligible. This simplified problem,
which we henceforth refer to as the `linear problem' is amenable to a
Green's function analysis which reveal several interesting physical
phenomena whose analogues are observed in the non-linear problem. Our
ultimate strategy will be to take approximations of the individual
contributions to the Green's function in the linear problem and
compare them with numerical simulations performed in the non-linear
regime to quantify the deviations. This paper deals with the linear
problem in horizon-penetrating coordinates compatible with NR.

\begin{figure}[t] 
	\centering
	\includegraphics[width=0.5\textwidth]{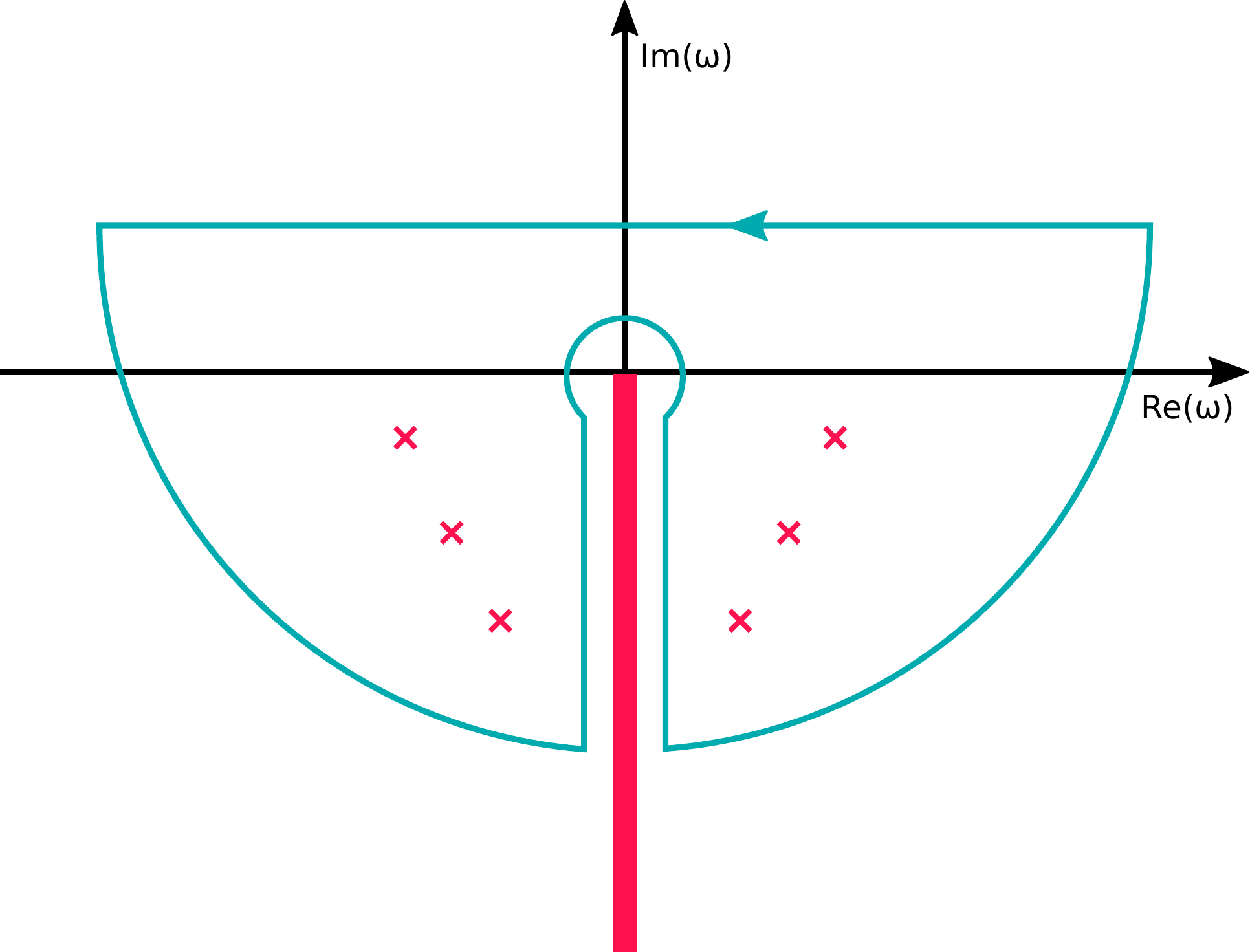}
	\caption{Singularities and branch cut of the Green's function
          in the~$\omega$ plane. The crosses denote singularities
          corresponding to the quasinormal mode frequencies while the
          magenta line indicates the branch cut along the negative
          imaginary $\omega$-axis. The contour of integration is represented by
          the blue curve.}
	\label{fig:GF}
\end{figure}

A scalar field signal, as seen by an observer outside the event
horizon, shows three generic features: a part from the direct
transmission of the initial data followed by QNM ringing and then a
tail which, at late times follows a power law. These features arise
from three different contributions to the Green's function: the high
frequency arc, the poles and the branch cut respectively. This is
depicted schematically in Fig.~\ref{fig:GF}. While calculating the QNM
contribution to the signal, we would like to obtain the dynamic
excitation amplitudes~\cite{Andersson1997} as opposed to assigning
constant excitation strengths to each QNM. This sidesteps the `timing
problem' which arises in the latter approach. The timing problem
essentially requires a choice of a starting time for observation such
that computed integrals do not diverge after that time. This turns out
to be problematic when the initial data is not sharply localized
because in that case the starting time is
ill-defined~\cite{Berti2006}.

The evolution of a massless scalar field is governed by the
Klein-Gordon equation,
\begin{align}
  \frac{1}{\sqrt{-g}}\frac{\partial}{\partial x^\mu}
  \left( \sqrt{-g} g^{\mu \nu} \frac{\partial \Phi}
  {\partial x^\nu}\right)=0,
  \label{eq:mKG}
\end{align}
where~$g_{\mu \nu}$, $g^{\mu \nu}$ and~$g$ are the components of the
metric, those of the inverse metric and the determinant of the metric
respectively. While our calculations would work also for complex
scalar fields, we will evolve real scalar fields presently. We
consider two sets of coordinates, the Kerr-Schild
coordinates~$\{t,r,\theta,\phi\}$ and generalized
coordinates~$\{T,r,\theta,\phi\}$ with the two time coordinates
related by
\begin{align}
t = T + h(r).
\end{align}
The height function~$h$ may be chosen arbitrarily but has a radial
asymptotic limit~$h \sim r$, $h' \sim 1$ near future-null infinity for
hyperboloidal slices, spacelike slices which terminate at future
null-infinity. In these coordinates, the line element for the
Schwarzschild metric can be written as
\begin{align}
  ds^2 &= - \left(1 - \frac{2M}{r}\right) dT^2
  + \left(\frac{4M}{r} + \frac{4 M h'}{r} - 2 h'\right) dT dr 
  \nonumber\\
  &+ \left(1 + \frac{2 M}{r} + \frac{4 M h'}{r} - h'^2
  + \frac{2 M h'^2}{r}\right) dr^2 \nonumber \\
  &+ r^2 (d\theta^2 + \sin^2 \theta d\phi^2).
\end{align}
The field is expanded in a basis of spherical harmonics according to
the ansatz,
\begin{align}
\Phi(T,r,\theta,\phi) = \sum_{l,m} K_{l,m}(T,r) Y_{l,m}(\theta,\phi).
\end{align}
The coefficients~$K_{l,m}$ are obtained using
\begin{align}
  K_{l,m} (T,r) =
  \int_{\theta = 0}^{\pi} \int_{\phi = 0}^{2 \pi}\Phi(T,r,\theta,\phi)
  Y^*_{l,m}(\theta,\phi) \sin \theta d\theta d\phi,
\end{align}
with~$*$ denoting the complex conjugate as usual. An initial
configuration of the scalar field is provided by
specifying~$K_{l,m}(0,r)$ and~$\left.\partial_T
K_{l,m}(T,r)\right|_{T=0}$ for every~$\left(l,m\right)$. The time
evolution of the scalar field can then be computed using the retarded
Green's function
\begin{align} 
K_{l,m}(T,r) &= \int G(T,r,r') \partial_T K_{l,m}(T,r')|_{T=0} dr'
\nonumber\\ + &\quad \int \partial_T G(T,r,r') K_{l,m}(0,r') dr'.
\label{eq:GF1st}
\end{align}
Our main objective throughout the rest of the analysis is to compute
the different parts of the retarded Green's function for the QNMs, the
tail and the direct transmission of the initial data. To ensure that
causality is respected, the above convolution is only performed over
the part of the initial data which lies within the past light cone of
the observer. To determine this, the coordinate light-speeds of the
left moving and right moving solutions must be computed from the roots
of the quadratic equation for~$v = dr/dT$
\begin{align}
  &\quad\left(1 + \frac{2 M}{r} + \frac{4 M h'}{r} - h'^2
  + \frac{2 M h'^2}{r}\right) v^2 \nonumber \\
  &+ \left(\frac{4M}{r} + \frac{4 M h'}{r} - 2 h'\right) v
  - \left(1 - \frac{2M}{r}\right) = 0. 
\end{align}
In Kerr-Schild coordinates, that is with~$T = t$, and thus~$h = 0$,
the upper limit of the integration is~$r' = r + T$ while the lower
limit is obtained by solving for~$r'$ in
\begin{align}
r' + 4 M \log(r' - 2M) = r + 4 M \log(r - 2 M) - T.
\end{align}
To reduce the wave equation into an ordinary differential equation, we
perform a Laplace transformation
\begin{align}
\hat{G}(\omega,r,r') = \int_{0}^{\infty} G(T,r,r') e^{i \omega T} dT,
\end{align}
with the inverse transform defined as
\begin{align} \label{eq:GFinverse}
G(T,r,r') = \frac{1}{2 \pi} \int_{-\infty + ic}^{\infty + ic}
\hat{G}(\omega,r,r') e^{-i \omega T} d \omega,
\end{align}
where~$c$ is some positive number.

The retarded Green's function in the frequency domain can then be
constructed from two linearly independent solutions to the ordinary
differential equation,
\begin{align}
&r (r-2 M) \frac{d^2 \hat{K}_{l,m}}{dr^2}
  + 2 \big(r - M - 2 i \omega M r \nonumber \\
  &+ i \omega r(r-2 M) h'\big) \frac{d \hat{K}_{l,m}}{dr} 
  +  \big(r \omega ^2 (r + 2M) -l (l+1) \nonumber \\
  &-  2 i \omega M + i  \omega r (r-2M) h''
  -r \omega ^2 (r-2 M) h'^2 \nonumber \\
  &+ 4 M r \omega ^2 h' + 2 i \omega  (r-M) h'\big)
\hat{K}_{l,m} = 0 , \label{eq:mainGFequation}
\end{align}
with each solution satisfying one of the boundary conditions for the
problem.

Introducing
\begin{align} \label{eq:Keqn}
  \hat{K}_{l,m} = e^{- i \omega (h \pm r)} H_{l,m}(r/2M),
\end{align}
and rescaling the coordinate according to~$x = r/ (2 M)$, we arrive at
the confluent Heun equation (CHE)~\cite{Fiziev1,Fiziev2,Fiziev3},
\begin{align}
&\frac{d^2}{dx^2}H_{l,m} (x) + \left( \alpha + \frac{\beta + 1}{x}
+ \frac{\gamma + 1}{x-1}\right)
\frac{d}{dx}H_{l,m}(x)\nonumber\\
&\qquad+ \left(\frac{\mu}{x} + \frac{\nu}{x-1}\right) H_{l,m}(x) = 0,
\label{eq:CHE}
\end{align}
with parameters independent of~$h(r)$, giving
\begin{align} 
  \alpha &= -4 i \bar{\omega},  && \beta = 0, &&
  \gamma = - 4 i \bar{\omega},\nonumber \\
  \mu &= l(l+1), && \nu = -l(l+1)
  - 4 i \bar{\omega}, \label{eq:CHEpar}
\end{align}
for the choice of minus sign in Eqn.~\eqref{eq:Keqn}. Here we have
defined~$\bar{\omega} = \omega M$. This is the form of the equation
that we will use for our calculations. An alternative form of the CHE
can be written for the plus sign in Eqn.~\eqref{eq:Keqn}, with
\begin{align}
\alpha &= 4 i \bar{\omega},  && \beta = 0,
&& \gamma = - 4 i \bar{\omega}, \nonumber \\
\mu &= l(l+1) + 4 i \bar{\omega}, && \nu = -l(l+1) + 16 \bar{\omega}^2. 
\end{align}

\subsection{The confluent Heun equation} \label{Section:HeunSolutions}

The Heun functions and its confluent forms have been used to describe
physical phenomenon in several disciplines of physics from quantum
mechanics and atomic physics to general relativity. A summary of
several important papers in physics is provided in~\cite{hortacsu}. In
black hole perturbation theory, some prominent applications of the CHE
include describing exact solutions of the Regge-Wheeler
equation~\cite{Fiziev1,Fiziev4}, wave equation in
Eddington-Finkelstein and Painleve-Gullstrand coordinates~\cite{DV},
the Teukolsky master equation for the Kerr-Neumann black
hole~\cite{VIEIRA201414, Fiziev2} and for describing the interior of
black hole spacetimes~\cite{FizievInterior1} among other things. The
solutions of the CHE has been expressed as a series solution of other
special functions in several interesting papers listed in the
references of~\cite{ishkhanyan}.

In this section, we briefly summarize local solutions of the CHE in
the existing literature and then write down asymptotic solutions in
terms of special functions of the confluent hypergeometric class. This
will be done for arbitrary parameters of the CHE and then for the
parameters pertaining to our problem, that is Eqn.~\eqref{eq:CHEpar}.

The CHE arises from the general Heun equation when two of its regular
singularities undergo a confluence to form an irregular
singularity. The CHE has five parameters and three singularities ---
two regular singularities at $x = 0,1$ and one irregular singularity
of rank $2$ at $x = \infty$~\cite{NIST:DLMF}. A summary of the
Frobenius and Thom{\'e} exponents are represented by its generalized
Riemann scheme (GRS)~\cite{SlavyanovBook} of our CHE given as
\begin{align}
\begin{pmatrix}
1 & 1 & 2 \\
0 & 1 & \infty \\
0 & 0 & \frac{\mu + \nu}{\alpha} & ; & x \\
-\beta & -\gamma & \beta + \gamma + 2 - \frac{\mu + \nu}{\alpha}\\
& & 0 \\
& & -\alpha   
\end{pmatrix}.
\end{align}
The GRS summarizes important information about the singularities and
the local solutions around those singularities. The first row
specifies the rank of the singularities and the second row specifies
their corresponding positions. The remaining rows specify the
Frobenius and Thom{\'e} exponents of the local solutions around these
singularities.
 
The canonical solution of the CHE is denoted by~$H_C(\alpha, \beta,
\gamma, \delta,\eta,x)$ where~\cite{Fiziev3}
\begin{align}
  \mu &= \frac{1}{2}(\alpha - \beta - \gamma + \alpha \beta
  - \beta \gamma) - \eta,\nonumber\\
  \nu &= \frac{1}{2}(\alpha + \beta + \gamma + \alpha \gamma
  + \beta \gamma) + \delta + \eta.
\end{align}
This solution is written as a convergent power series about the origin,
\begin{align}
  H_0^{(1)} = H_C(\alpha, \beta, \gamma, \delta, \eta,x)
  = \sum_{n = 0}^{\infty} a_n x^n, && |x|<1,
\end{align}
with coefficients satisfying a three-term recurrence relation,
\begin{align}
  \alpha_n a_{n+1} + \beta_n a_{n} + \gamma_n a_{n-1} = 0, \label{eq:recurrence}
\end{align}
where~$a_{-1} = 0, \ a_{0} = 1$ and
\begin{align}
\alpha_n &= -n^2 - n(\beta +2) -1-\beta, \nonumber \\
\beta_n &= n^2+n (-\alpha +\beta +\gamma +1) -\mu, \nonumber \\
\gamma_n &= n \alpha + (\mu + \nu - \alpha).
\end{align}
The second solution can be written in terms of this canonical solution
as~\cite{DV}
\begin{align}
	H_{0}^{(2)} = x^{-\beta} H_C(\alpha, -\beta, \gamma, \delta, \eta,x).
\end{align}
Similarly, two local Frobenius solutions can be constructed
about~$x=1$ which can be written in terms of the canonical solution as
\begin{align}
H^{(1)}_1 &= H_C(-\alpha,\gamma,\beta,-\delta,\eta+\delta,1-x), \nonumber\\
H^{(2)}_1 &= (x-1)^{-\gamma}H_C(-\alpha,-\gamma,\beta,-\delta,\eta+\delta,1-x).
\end{align}
The first of this pair is of interest to us as this solution has the
desired behavior of a QNM near the horizon. However, since this
solution converges within a unit circle centered at~$x = 1$, it must
be analytically continued to cover the entire positive $x$-axis. This
shall be discussed in some detail in section~\ref{subsection:QNMs}.

Following~\cite{Olver}, we can write down two asymptotic solutions in
the vicinity of the irregular singular point in a power series
of~$1/x$
\begin{align}
  H^{(1)}_{\infty} &\approx x^{-\frac{\mu+\nu}{\alpha}}
  \sum_{n=0}^{\infty} \frac{a_n}{x^n},
  \nonumber\\
  H^{(2)}_{\infty} &\approx e^{-\alpha x} x^{-\beta - \gamma -
    2 + (\mu + \nu)/\alpha} \sum_{n=0}^{\infty} \frac{a_n}{x^n}.
\end{align}
It must be noted here that these Thom{\'e} solutions may not
necessarily converge. The coefficients~$a_n$ can be calculated using
the recurrence relation
\begin{align}
(\alpha + 2 p) n a_n &= (n - q - \beta - \gamma - 2)(n - 1 - q) a_{n-1}.
\end{align}
Here~$a_0$ and~$a_1$ are arbitrary and~$p$, $q$ are constructed from
the Thom{\'e} exponents with~$p = 0,-\alpha$ and $q = -(\mu +
\nu)/\alpha, -\beta - \gamma - 2 + (\mu + \nu)/\alpha$ for the two
solutions.

It is also possible to alternatively represent asymptotic solutions of
the CHE using special functions. First, the CHE must be converted into
the normal form which removes the first derivative using the
transformation
\begin{align}
U(x) = e^{\frac{1}{2} \alpha x} (x-1)^{\frac{1}{2}(1+\gamma)} x^{\frac{1}{2}(1+\beta)} H(x).
\end{align}
$U(x)$ then satisfies the differential equation
\begin{align}
\frac{d^2 U}{dx^2} + R U = 0,
\end{align}
with
\begin{align}
R &= \frac{1}{2} \left(\frac{\beta +1}{x^2}+\frac{\gamma +1}{(x-1)^2}
\right)-\frac{1}{4} \left(\alpha +\frac{\beta +1}{x}
+\frac{\gamma +1}{x-1}\right)^2 \nonumber \\
&\quad+\frac{\mu }{x}+\frac{\nu }{x-1}.
\end{align}
Expanding~$R$ in powers of~$1/x$, we can obtain several asymptotic
forms of the above equation depending on the power of~$1/x$ at which
we truncate~$R$. To begin with, we neglect~$\mathcal{O}(1/x^2)$ and
higher order terms to arrive at
\begin{align} 
\frac{d^2 U}{dx^2}+\left(-\frac{\alpha ^2}{4} + \frac{-\frac{\alpha  \beta }{2}
  -\frac{\alpha  \gamma }{2}-\alpha +\mu +\nu }{x}\right) U \approx 0,
\label{eq:Whittaker1}
\end{align}
which is a Whittaker equation and has the standard Whittaker
functions~$M_{a,b}$, $W_{a,b}$ as solutions, the definitions of which
are provided in~\cite{NIST:DLMF}. Alternatively, the Tricomi and
Kummer confluent hypergeometric functions can also be used as
solutions, using their relations with the Whittaker functions. In
terms of~$M_{a,b}$ and~$W_{a,b}$, the solutions take the form
\begin{align}
H^{(1)}_{\infty} \approx \frac{M_{\frac{2 (\mu +\nu )-\alpha
(\beta +\gamma +2)}{2 \alpha },\frac{1}{2}}(\alpha x)}{e^{\frac{1}{2} \alpha x}
(x-1)^{\frac{1}{2}(1+\gamma)} x^{\frac{1}{2}(1+\beta)}}, \nonumber\\ 
H^{(2)}_{\infty} \approx \frac{W_{\frac{2 (\mu +\nu )-\alpha
(\beta +\gamma +2)}{2 \alpha },
\frac{1}{2}}(\alpha x)}{e^{\frac{1}{2} \alpha x} (x-1)^{\frac{1}{2}(1+\gamma)}
x^{\frac{1}{2}(1+\beta)}}. \label{eq:WhittSoln12}
\end{align}
Another asymptotic form of the
solutions can be obtained when we neglect~$\mathcal{O}(1/x^3)$ and
higher order terms, which leads to
\begin{widetext}
\begin{align} \label{eq:Whittaker2}
\frac{d^2 U}{dx^2}+\left(-\frac{\alpha ^2}{4}+\frac{-\frac{1}{2}
\alpha  (\beta +\gamma +2)+\mu +\nu }{x} +\frac{\frac{1}{4}
\left(-2 \alpha  (\gamma +1)-(\beta +\gamma +2)^2\right)+\frac{1}{2}
(\beta +\gamma +2)+\nu }{x^2}\right) U \approx 0,
\end{align}
\end{widetext}
This is also a Whittaker equation and its solutions are given by
\begin{align} 
H^{(1)}_{\infty} \approx \frac{M_{\frac{2 (\mu +\nu )-\alpha
(\beta +\gamma +2)}{2 \alpha },\frac{1}{2} \sqrt{(\beta +\gamma +1)^2+2
\alpha  (\gamma +1)-4 \nu }}(\alpha x )}{e^{\frac{1}{2} \alpha x}
  (x-1)^{\frac{1}{2}(1+\gamma)} x^{\frac{1}{2}(1+\beta)}},\label{eq:WhittSoln34a}
  \nonumber \\ 
\end{align} 
and
\begin{align} 
H^{(2)}_{\infty} \approx \frac{W_{\frac{2 (\mu +\nu )-\alpha
(\beta +\gamma +2)}{2 \alpha },\frac{1}{2} \sqrt{(\beta +\gamma +1)^2
+2 \alpha  (\gamma +1)-4 \nu }}(\alpha x )}{e^{\frac{1}{2} \alpha x}
(x-1)^{\frac{1}{2}(1+\gamma)} x^{\frac{1}{2}(1+\beta)}}. \label{eq:WhittSoln34b}
\end{align}
Using asymptotic forms of the Whittaker functions, we can show that
the two sets of asymptotic forms Eqns.~\eqref{eq:WhittSoln12}
and~\eqref{eq:WhittSoln34a}-\eqref{eq:WhittSoln34b} exhibit the same
behavior when~$x \rightarrow \infty$, namely
\begin{align}
H^{(1)}_{\infty} &\approx x^{-\frac{\mu + \nu}{\alpha}}, && |\arg(\alpha)|
\leq \frac{1}{2}\pi, \nonumber\\  
H^{(2)}_{\infty} &\approx e^{-\alpha x} x^{\frac{\mu + \nu}{\alpha}}
x^{-(2 + \beta + \gamma)}, && |\arg(\alpha)| \leq \frac{3}{2}\pi.
\end{align}
Now with the specific choice of parameters specified in
Eqn.~\eqref{eq:CHEpar}, the two parameters in the alternative notation
are given by
\begin{align} \eta = -l(l+1), && \delta = 8
      \bar{\omega}^2.
\end{align}
The GRS of our CHE can then be written
\begin{align} \label{eq:GRSproblem}
\begin{pmatrix}
1 & 1 & 2 \\
0 & 1 & \infty \\
0 & 0 & 1 & ; & x \\
0 & 4 i \bar{\omega} & 1 - 4 i \bar{\omega}\\
& & 0 \\
& & 4 i \bar{\omega}   
\end{pmatrix}.
\end{align}
Using the GRS, we can write the two Frobenius solutions about~$x=1$
and the two Thom{\'e} solutions in a straightforward manner,
\begin{align}
  H_1^{(1)} &= H_C(4i \bar{\omega},-4i
  \bar{\omega},0, -8 \bar{\omega}^2,8 \bar{\omega}^2-l(l+1),1-x),
  \nonumber\\
  H_1^{(2)} &= (x-1)^{4 i \bar{\omega} }\times\nonumber\\
  &\quad H_C(4i\bar{\omega}, 4i \bar{\omega},0,-8 \bar{\omega}^2,8
  \bar{\omega}^2-l(l+1),1-x), \nonumber \\
  H_{\infty}^{(1)} &\approx
  x^{-1} \sum_{n=0}^{\infty} \frac{a_n}{x^n},
  \quad H_{\infty}^{(2)} \approx e^{4 i \bar{\omega} x} x^{-1
    + 4 i \bar{\omega}} \sum_{n=0}^{\infty}
  \frac{a_n}{x^n}. \label{eq:ThomeSoln}
\end{align}
The alternative representations of the asymptotic solutions in terms
of the Whittaker functions can then be computed assuming~$1\ll x$,
giving
\begin{align}
H_{\infty}^{(1)} &\approx \frac{e^{2 i \bar{\omega} x}
(x-1)^{-\frac{1}{2}+2 i \bar{\omega} }}{\sqrt{x}}
M_{2 i \bar{\omega} ,\frac{1}{2} \sqrt{(2 l+1)^2-48 \bar{\omega}^2}}
(-4 i \bar{\omega} x ), \nonumber\\
H_{\infty}^{(2)} &\approx \frac{e^{2 i \bar{\omega} x}
 (x-1)^{-\frac{1}{2}+2 i \bar{\omega} }}
{\sqrt{x}} W_{2 i \bar{\omega} ,\frac{1}{2} \sqrt{(2 l+1)^2-48 \bar{\omega}^2}}
(-4 i \bar{\omega} x ),
\end{align}
or alternatively
\begin{align}
H_{\infty}^{(1)} &\approx \frac{e^{2 i \bar{\omega} x}
(x-1)^{-\frac{1}{2}+2 i \bar{\omega} }}{\sqrt{x}} M_{2 i \bar{\omega},
\frac{1}{2}}(-4 i \bar{\omega} x ), \nonumber\\
H_{\infty}^{(2)} &\approx \frac{e^{2 i \bar{\omega} x} (x-1)^{-\frac{1}{2}
+2 i \bar{\omega} }}{\sqrt{x}} W_{2 i \bar{\omega} ,\frac{1}{2}}(-4 i \bar{\omega} x ).
\end{align}
These solutions are only valid in the vicinity of the irregular
singular point and can be expressed by the limiting forms as~$x
\rightarrow \infty$,
\begin{align}
H_{\infty}^{(1)} &\approx  x^{-1}, \qquad
H_{\infty}^{(2)} \approx  e^{4 i \bar{\omega} x} x^{-1 + 4 i \bar{\omega}}.
\end{align}

\subsection{Quasinormal modes}
\label{subsection:QNMs}        

\subsubsection{QNM boundary conditions}

Quasinormal modes are solutions of the eigenvalue problem of the
Regge-Wheeler equation~\cite{RW} with purely outgoing boundary
conditions at the horizon and at spatial infinity. The QNM
frequencies, which are complex, correspond to the poles of the Green's
function to the wave equation, and as we shall see, are frequencies at
which the Wronskian of the two linearly independent solutions used to
construct the Green's function vanishes. The outgoing boundary
conditions, when applied to~$H$ take the form
\begin{align}
H &\sim \frac{1}{2Mx}, && x \rightarrow 1,\nonumber\\
H &\sim \frac{1}{2M} e^{4 i \bar{\omega} x} x^{-1 + 4  i \bar{\omega}},
&& x \rightarrow \infty.
\end{align}
Here we have chosen the normalization constants such that these
conditions are identical to their counterparts in Regge-Wheeler
coordinates in the literature~\cite{Nollert, BertiReview}. In the
original coordinates~$\{T,r,\theta,\phi\}$, they become
\begin{align}
K &\sim \frac{1}{r} e^{-i \omega (r+h)}, && r \rightarrow 2 M,\nonumber\\
K &\sim \frac{1}{r} e^{i \omega (r-h)} \left(\frac{r}{2 M}\right)^{4 i \bar{\omega}},
&& r \rightarrow \infty.
\end{align}
The first of these implies that QNM solutions are finite at the future
horizon. This feature is explicit in our treatment because of the use
of horizon penetrating coordinates. Although it is most convenient to
construct QNM solutions in the standard Schwarzschild time coordinate
with that choice the solutions appear irregular at the horizon. This
is misleading, because the blow-up occurs at the bifurcation sphere
where the Schwarzschild foliation meets the horizon, but not
elsewhere. Pure QNM data can thus can be evolved with standard
numerical relativity tools, provided the outer boundary is treated
appropriately. This can be seen clearly in Kerr-Schild coordinates by
setting~$h(r)=0$. The problems at the outer boundary can be ideally
avoided by employing a hyperboloidal foliation~\cite{ansorgrodrigo},
which we will employ in future work. Choosing a suitable height
function~$h(r) = r + 4 M \log r$, both boundary conditions are
regular,
\begin{align}
  K &\sim \frac{1}{r} e^{- 2 i \omega r} r^{-4 i \bar{\omega}}, &&
  r \rightarrow 2 M,\nonumber\\
K &\sim \frac{1}{r} \left(\frac{1}{2 M}\right)^{4 i \bar{\omega}},
&& r \rightarrow \infty.
\end{align}
We will now discuss several solutions of the CHE which satisfy at
least one of these boundary conditions, and then construct global
solutions which satisfy both boundary conditions simultaneously but
only at the QNM frequencies.

\subsubsection{Solution satisfying boundary condition at the horizon~($f_{-}$)}

The Frobenius solution around~$x = 1$ which is bounded satisfies the
boundary condition at the horizon. This solution can be written in
terms of the canonical solution of the CHE with appropriate
normalization
\begin{align} \label{eq:FrobeniusSoln}
  H = \frac{1}{2M} H_C(4i \bar{\omega},-4i\bar{\omega},
  0,-8\bar{\omega}^2,8 \bar{\omega}^2-l-l^2,1-x).
\end{align}
This solution converges between~$0 < x < 2$ but can be analytically
continued to converge over the entire positive~$r$ axis. This will be
discussed later in the section.

Another solution of importance which satisfies the same boundary
conditions is a convergent series solution in terms of the Gauss
hypergeometric functions following the lines of Mano, Suzuki and
Tagasuki (MST)~\cite{MSTa,MSTb,Casals1},
\begin{align}
  H &= \frac{1}{N_{F}}\sum_{n=-\infty}^{\infty} a_n
  \frac{\Gamma (-n-\nu -2 i \bar{\omega} )
    \Gamma (n+\nu -2 i \bar{\omega} +1)}{\Gamma (1-4 i \bar{\omega} )}  \\
  & \times {}_2F_1(-\nu -n-2 i \bar{\omega} ,\nu
  +n-2 i \bar{\omega} +1;1-4 i \bar{\omega} ;1-x), \nonumber
\end{align}
where~${}_2F_1$ is the Gauss hypergeometric function, $a_0$ is equal
to~$1$ and the normalization condition is given by
\begin{align}
  N_F = 2 M\sum_{n=-\infty}^{\infty} a_n \frac{\Gamma (-n-\nu -2 i \bar{\omega} )
    \Gamma (n+\nu -2 i \bar{\omega} +1)}{\Gamma (1-4 i \bar{\omega} )}.
\end{align}
The coefficients~$a_n$ satisfy a three-term recurrence relation as in
Eqn.~\eqref{eq:recurrence} with
\begin{align}
  \alpha_n &= - \frac{2 i \bar{\omega} (n + \nu + 1 - 2 i \bar{\omega})
    (n + \nu + 1 - 2 i \bar{\omega})}{(2n + 2 \nu + 3)} \nonumber \\
  &\quad\times (n + \nu + 1 + 2 i \bar{\omega})(n + \nu), \nonumber \\
  \beta_n &= - l(l+1) (n + \nu)(n + \nu + 1) \nonumber \\
  &\quad+ ((n + \nu)(n + \nu + 1) + 4 \bar{\omega}^2)^2, \nonumber \\
  \gamma_n &= \frac{2 i \bar{\omega}(n + \nu + 2 i \bar{\omega})^2
    (n + \nu - 2 i \bar{\omega})(n + \nu + 1)}{(2n + 2 \nu - 1)}.
\end{align}
The parameter~$\nu$ called the renormalized angular momentum is
determined by the fact that the series should converge both as $n
\rightarrow \infty$ and $n \rightarrow - \infty$. This is ensured by
solving the transcendental equation~\cite{MSTa, MSTb}:
\begin{align}
P_n(\nu) Q_{n-1}(\nu) = 1,
\end{align}
where the continued fractions are given by:
\begin{align}
P_n(\nu) &= \frac{a_n}{a_{n-1}}, && Q_n(\nu) = \frac{a_n}{a_{n+1}}, \\
P_n(\nu) &= - \frac{\gamma_n}{\beta_n + a_n P_{n+1}(\nu)}, && Q_n(\nu)
= - \frac{\alpha_n}{\beta_n + a_n Q_{n-1}(\nu)}. \nonumber
\end{align}
When the renormalization parameter is chosen correctly, the series
solution converges between $1 < x < \infty$.

\subsubsection{Solution satisfying boundary condition at infinity ($f_{+}$)}

Solutions satisfying the boundary condition at infinity can be
constructed from Whittaker functions or equivalently the confluent
hypergeometric functions following the lines of Leaver's U-series
solutions~\cite{Leaver2}
\begin{align}
H = \frac{ e^{4 i \bar{\omega} x}}{N_U} \sum_{n=0}^{\infty}
a_n \Gamma(1 + n - 4 i \bar{\omega})
U(1 + n - 4 i \bar{\omega},1, - 4 i \bar{\omega} x),
\end{align}
with~$a_0=1$ and the normalization constant
\begin{align}
N_U = 2 M (- 4 i \bar{\omega})^{-1 + 4 i \bar{\omega}}
\Gamma(1 - 4 i \bar{\omega}).
\end{align}
The coefficients~$a_n$ also satisfy a three-term recurrence relation
as in Eqn.~\eqref{eq:recurrence}, which match with the recurrence
relations of Leaver's Jaff{\'e} series~\cite{Leaver1, Leaver2}
\begin{align} \label{eq:Jafferecurrence}
\alpha_n &= 1 - 4 i \bar{\omega} + (2 - 4 i \bar{\omega}) n + n^2, \nonumber \\
\beta_n &= 32 \bar{\omega}^2 + 8 i \bar{\omega} - 1 - l(l+1)
+ (16 i \bar{\omega} - 2) n - 2 n^2, \nonumber \\
\gamma_n &= - 16 \bar{\omega}^2 - 8 i \bar{\omega} n + n^2.
\end{align}
This solution is uniformly convergent as~$x \rightarrow \infty$ and
diverges as~$x \rightarrow 1$ when~$\bar{\omega}$ is not an
eigenfrequency. It is absolutely convergent on any interval bounded
away from~$x=1$~\cite{Leaver2}.

\subsubsection{Solution satisfying both boundary conditions}
\label{subsubsection:exactsoln}

To construct solutions which satisfy both boundary conditions
simultaneously, we have to solve the central two-point connection
problem for the CHE which connects local solutions with the desired
behavior at the two endpoints of an interval. This problem, in its
most general form requires the construction of a connection matrix
binding these local solutions and at present remains unsolved for the
Heun class of differential equations. Hence, we only look at
eigenvalues at which both boundary conditions are satisfied.

A detailed description of the method is provided
in~\cite{SlavyanovBook} and~\cite{DV}, so only the approach is
outlined here,
\begin{enumerate}[I.]
\item The local solutions at the horizon are to be connected with
  those at spatial infinity. Hence we shift the singularities at~$0$
  and~$1$ to~$-1$ and~$0$ respectively,
\begin{align}
x \longmapsto z = x - 1, && H(x) \rightarrow S(z).
\end{align}
\item The next step is to perform an s-homotopic transformation which
  makes the solution around~$z=0$ bounded for arbitrary values of the
  eigenvalue~$\omega$ while the asymptotic behavior at infinity is
  given by a linear combination of the two Thom{\'e}
  solutions in Eqn.~\eqref{eq:ThomeSoln},
\begin{align}
S(z) = e^{4 i \bar{\omega} z} (z+1)^{-1 + 4 i \bar{\omega}} T(z).
\end{align}
\item Finally, a M{\"o}bius transformation brings the irregular
  singularity to~$y=1$ while the position of the singularity at the
  origin remains unchanged,
\begin{align}
z \longmapsto y = \frac{z}{z+1}, && T(z) \rightarrow U(y).
\end{align}	

\end{enumerate}
After these two transformations, which are together referred to as the
Jaff{\'e} transformation, we obtain the following ODE:
\begin{align}
  &y(y-1)^2 U'' + \big(1 - 4 i \bar{\omega}
  + (16 i \bar{\omega} - 4) y
  + (3 - 8 i \bar{\omega}) y^2\big) U' \nonumber\\
  &+ \big(8 i \bar{\omega} + 32 \bar{\omega}^2 -1 - l(l+1)
  + (1 - 8 i \bar{\omega} - 16 \bar{\omega}^2) y \big)U = 0.
\end{align}
Now the eigenvalue problem is to be solved between~$\left[0,1\right]$
in~$y$ and there are no other singularities in that interval. A
Jaff{\'e} expansion, which is a power-series expansion of the form
\begin{align}
	U(y) = \sum_{n=0}^{\infty} a_n y^n,
\end{align}
is always convergent in the unit circle about~$y = 0$. In the original
coordinates, this results in a solution which is convergent in~$1/2 <
x < \infty$
\begin{align}
H = \frac{1}{2M e^{4 i \bar{\omega}}} e^{4 i \bar{\omega} x}
x^{-1 + 4 i \bar{\omega}} \sum_{n=0}^{\infty} a_n \left(\frac{x - 1}{x}\right)^n, 
\end{align}
where the coefficients~$a_n$ follow a three term recurrence relation
as in Eqn.~\eqref{eq:recurrence} with coefficients matching those of
Leaver's Jaffe series as in Eqn.~\eqref{eq:Jafferecurrence}. This
solution coincides with the desired Frobenius solution at~$x=1$ in the
region of overlap and can therefore be used to construct a
representation of the confluent Heun function which is convergent
in~$0 < x < \infty$.

The boundary conditions at spatial infinity are only satisfied
when~$\sum a_n$ is finite, that is the series is absolutely
convergent. This only holds true for specific values of the complex
frequency which can be found out by solving the continued fraction
equation for~$\omega$
\begin{align}
  0 = \beta_0 - \frac{\alpha_0 \gamma_1}
  {\beta_1 -\frac{\alpha_1 \gamma_2}{\beta_2 - \ldots}}
\label{eq:CFE}
\end{align}
Using the recurrence relations from Eqn.~\eqref{eq:Jafferecurrence},
this equation is identical to that of Leaver~\cite{Leaver1} and hence
results in the same frequencies. An alternative method to obtain QNM
frequencies using the CHE is provided in~\cite{FizievQNMsPRD}.

\subsection{The exact Green's function}
\label{subsection:Exact_Green}

The differential operator in question is a non self-adjoint,
non-Hermitian operator whose Green's function satisfies the following
differential equation, now reverting to our `physical'
coordinates~$(T,r,\theta,\phi)$
\begin{align}
    &\frac{d}{dr} \left( w(\omega,r)
    \frac{d \hat{G}(\omega,r,r')}{dr}\right) + V(\omega,r)
    \hat{G}(\omega,r,r')\nonumber\\
&= p(\omega,r)\delta(r-r'),
\end{align}
where
\begin{align}
  p &= e^{2 i \omega h} r^2 (r-2 M)^{-4 i \omega M}, \nonumber \\
  w &= e^{2 i \omega h} r (r-2 M)^{1-4 i \omega M}, \nonumber \\
  V &=  e^{2 i \omega h} (r-2 M)^{-4 i \omega M} \big(r \omega ^2 (r + 2M) \nonumber \\
  &+\omega  h' \left(4 M r \omega - r \omega h'(r - 2 M)
  +2 i (r-M)\right) \nonumber \\
  &- 2 i \omega M -l (l+1) + i \omega r  (r-2M) h''\big).
\end{align}
The explicit form of the Green's function can be written down from the
two linearly independent solutions~$f_{-}$, $f_{+}$ of
Eqn.~\eqref{eq:mainGFequation} satisfying one of the boundary
conditions each, 
\begin{align}
\hat{G}(\omega, r, r') = \frac{1}{A(\omega)} \begin{cases}
p(\omega, r') f_{-}(\omega,r)f_{+}(\omega,r'), \ \ \ r \leq r', \\
p(\omega, r') f_{-}(\omega,r')f_{+}(\omega,r), \ \ \ r' < r.
\end{cases}
\end{align}
Here~$A(\omega)$ is the standard weighted Wronskian of the two
solutions
\begin{align}
A(\omega) = w(r) \left(f_{-} f'_{+} - f'_{-} f_{+} \right).
\end{align}
The Green's function has poles in the lower half of the~$\omega$-plane
and a branch cut along the negative imaginary $\omega$-axis, as shown
in Fig.~\eqref{fig:GF}. At the poles the~$f_{-}$ and~$f_{+}$ solutions
becomes proportional to the other and the weighted Wronskian
vanishes. The frequencies at which this happens are the QNM
frequencies computed from the continuous fraction equation,
Eqn.~\eqref{eq:CFE}. The contribution from the branch cut gives a
measure of the backscattering, which at late times generates a power
law decay. The two solutions~$f_-$ and~$f_+$ are
\begin{align}
f_{-} &= \frac{1}{2M e^{4 i \bar{\omega}}} e^{i \omega (r-h)}
\left(\frac{r}{2M}\right)^{-1 + 4 i \bar{\omega}}
\sum_{n=0}^{\infty} a_n \left(\frac{r - 2M}{r}\right)^n, \\
f_{+} &= \frac{ e^{ i\omega (r-h)}}{2 M (- 4 i \bar{\omega})^{-1 + 4 i \bar{\omega}}
\Gamma(1 - 4 i \bar{\omega})} \nonumber \\
& \times \sum_{n=0}^{\infty} a_n \Gamma(1 + n - 4 i \bar{\omega})
U(1 + n - 4 i \bar{\omega},1, - 2 i \omega r).
\end{align}
Note here that the presence of the arbitrary height-function allows us
to take care, within our analysis, of any spherically symmetric
foliation compatible with the timelike killing vector of the
background.

\subsection{Quasinormal mode excitation factors}\label{subsection:QNEF}

It is well known that in some region of spacetime, the solution to the
wave equation may be represented as a linear combination of spatially
truncated QNMs~\cite{szpak}. This can be seen when we construct the
part of the Green's function which encodes the contribution from the
poles. In doing so, as elsewhere, the poles are assumed to be simple,
that is, near the QNM frequency~$\omega_{l,n}$ the weighted Wronskian
has the form
\begin{align}
A(\omega_{l,n}) \approx (\omega - \omega_{l,n}) A'(\omega_{l,n}).
\end{align}
Using Eqn.~\eqref{eq:GFinverse}, the QNM part of the time domain
Green's function is given by
\begin{align}
G^{Q}(T,r,r') = \frac{1}{2 \pi} \sum_{l,n} \oint_{\omega_{l,n}}
\frac{p  \ f_{-} f_{+}}{(\omega - \omega_{l,n}) A'(\omega)} e^{-i \omega T} d \omega,
\end{align}
where dependence on~$r$ and~$r'$ has been suppressed for brevity.

Using the fact that the QNM frequencies are located symmetrically
about the negative imaginary $\omega$-axis, this integral can now be
easily solved by using Cauchy's residual theorem, giving
\begin{align}
  &G^{Q}(T, r, r') =  \sum_{l=0}^{\infty}\sum_{n=0}^{\infty}
  \frac{2 i e^{-i \omega_{l,n} T}}{A'(\omega_{l,n})}\times \nonumber\\
  &\quad\,\, \begin{cases}
  p(\omega, r') f_{-}(\omega_{l,n},r)f_{+}(\omega_{l,n},r'), \ \ \ r \leq r', \\
  p(\omega, r') f_{-}(\omega_{l,n},r')f_{+}(\omega_{l,n},r), \ \ \ r' \leq r. 
\end{cases} \label{eq:QNM}
\end{align}
This is the key formula in this section and can be used to calculate
the QNM contribution to the scalar field signal for any observer
outside the event horizon. This equation can be further simplified for
an asymptotic observer,~$r \rightarrow \infty$ by assuming that the
initial data has no support outside the observer, that is for~$r >
r'$,
\begin{align}
G^{Q}(T, r, r') &= \sum_{l,n}^{\infty}
\frac{2 i }{A'(\omega_{l,n})}
p(\omega ,r') f_{-}(\omega_{l,n},r')\times \nonumber \\
&\quad
\frac{1}{r} \left(\frac{r}{2 M}\right)^{4 i \omega_{l,n} M}
e^{-i \omega_{l,n} \left(T-r + h(r)\right)}.
\end{align}
The quantities~$B_{l,n}=2i/A'_{l,n}$ are called the quasinormal mode
excitation factors (QNEFs). A list of some of them can be found in
Table~\ref{tab:lineardata}. One point to note while
calculating~$A'_{l,n}$ is that as a control for its accuracy we check
the Cauchy-Riemann conditions with respect to~$\omega$ at the poles,
and keep the digits up-to which they are satisfied.

\begin{table*}[th!]
	\centering
	\begin{tabular}{ccccc}
		$l$ & $n$ & $\omega_{l,n}$ & $A'(\omega_{l,n})$ & $B_{l,n}$\\ 
		\hline \hline
		0 & 0 & $0.11045493908041968588-0.10489571708688095878i$ & $1.32962 + 3.01240i$ & $0.55567+0.24526i$\\
		& 1 & $0.08611691833639926-0.34805244680646047i$& $4.37158 + 0.92283i$ &  $0.09246+0.43798i$\\
		& 2 & $0.07574193553517584-0.6010785900358036i$ & $4.1171 + 0.1769i$ & $0.0208+0.4849i$\\
		& 3 & $0.0704101384174665-0.853677318105532i$ & $3.0109-0.02768i$ &  $-0.0061+0.66420i$\\
		& 4 & $0.0670743042285181-1.1056318799366185i$ & $1.97387-0.06594i$ &  $-0.0338+1.01211i$\\
		\hline
		1 & 0 & $0.29293613326728270862-0.097659988913578222156i$ & $-4.2778 + 3.3416i$ &  $0.2268-0.2904i$\\
		& 1 & $0.26444865060483253963-0.30625739155904712323i$ & $1.4742 + 2.4857i$ &  $0.5952+0.3530i$\\
		& 2 & $0.22953933493130167185-0.54013342501910721347i$ & $2.4729 + 0.70097i$ &  $0.2122+0.74862i$\\
		& 3 & $0.2032583861834636453-0.7882978227811980306i$ & $2.1992 - 0.08555i$ &  $-0.0353+0.90805i$\\
		& 4 & $0.185109020345202-1.040762112817569i$ & $1.6283 - 0.2783i$ &  $-0.2040+1.1934i$\\
		\hline
		2 & 0 & $0.48364387221071298673-0.096758775978287862659i$ & $-5.9991 - 3.6075i$ &  $-0.1472-0.2448i$\\
		& 1 & $0.46385057901976556322-0.29560393698796252621i$ & $-1.2614 + 1.7966i$ &  $0.7456-0.5235i$\\
		& 2 & $0.43054405437657576811-0.50855840215427448747i$ & $0.62862 + 1.1773i$ &  $1.32192+0.7058i$\\
		& 3 & $0.39386306288868911970-0.73809658478099752579i$ & $0.96947 + 0.3802i$ &  $0.70120+1.7880i$\\
		& 4 & $0.36129919188736593055-0.97992151947121679169i$ & $0.83335 - 0.02526i$ & $-0.07268+2.39775i$\\
		\hline
		3 & 0 & $0.67536623253662053532-0.09649962773400958388i$ & $1.2979 - 8.1785i$ &  $-0.2385+0.0379i$\\
		& 1 & $0.66067149795596247482-0.29228478513841188658i$ & $-1.8081 - 0.3815i$ &  $-0.2234-1.0590i$\\
		& 2 & $0.63362580769432366407-0.49600823040312675197i$ & $-0.41392 + 0.70464i$ &  $2.11018-1.23956i$\\
		& 3 & $0.59877325279995979383-0.71122120737134861358i$ & $0.2123 + 0.46522i$ &  $3.5580+1.62370i$\\
		& 4 & $0.56162728989869021279-0.93859282364463198356i$ & $0.33704 + 0.16698i$ &  $2.36050+4.76455i$\\
		\hline
		4 & 0 & $0.86741564173787901722-0.09639169234802256387i$ & $9.137 - 2.15913i$ &  $-0.049+0.20731i$\\
		& 1 & $0.85580803512377558870-0.29087602253327418949i$ & $-0.31151 -1.59767i$ &  $-1.20598-0.23514i$\\
		& 2 & $0.83369213256148927756-0.49032489461814627949i$ & $-0.561361 - 0.030657i$ &  $-0.193991-3.552176i$\\
		& 3 & $0.80328811286099866551-0.69748155123442989656i$ & $-0.1409823 + 0.245928i$ &  $6.1209168-3.508917i$\\
		& 4 & $0.76773262396440926056-0.91401943246331559159i$ & $0.063191 + 0.167472i$ &  $10.453941+3.944510i$\\
		\hline
	\end{tabular}
	\caption{Excitation factors for the Schwarzschild black hole
          with~$M=1$. The columns from left to right are: mode
          number~$l$, overtone number~$n$, QNM
          frequency~$\omega_{l,n}$ computed from the continuous
          fraction equation, derivative of the weighted Wronskian~$A'$
          evaluated at the QNM frequencies and the QNM excitation
          factor~$B_{l,n}$. \label{tab:lineardata}}
\end{table*}

Returning to the general case, the QNM response to some given initial
data can now be evaluated as
\begin{align} \label{eq:qnmsum}
K_{l,m}(T,r) = \sum_{n} C_{l,m,n} e^{-i \omega_{l,n} T},
\end{align}
where the QNM excitation amplitude~$C_{l,m,n}$ is given by
\begin{align}
C_{l,m,n} &= B_{l,n} \int p( r') f_-(r) f_{+}(r')
\partial_T K_{l,m}(T,r')|_{T=0} dr'  \nonumber \\
&\quad- i \omega_{l,n} B_{l,n} \int p(r') f_-(r) f_{+}(r') K_{l,m}(0,r') dr',
\end{align}
Here we have suppressed the fact that all functions are
evaluated at the QNM frequencies~$\omega_{l,n}$. As has been mentioned
before, the limits of this integration are functions of time and
therefore~$C_{l,m,n}$ are referred to as `dynamic' excitation
amplitudes~\cite{Andersson1997}. It is only meaningful to represent
solutions of the wave equation as a linear combination of QNMs in the
region which lies in the future light cone of the entire initial data,
which is also where these excitation amplitudes become
time-independent~\cite{szpak}.

\subsection{Tail rates}\label{subsection:Tails}

We now proceed to calculate the part of the Green's function which
encodes the contribution of the branch cut to the signal, the general
expression for which can be written down as~\cite{LeaverPRD,
  Andersson1997}
\begin{align} \label{eq:BCgeneral}
G^{B}(T,r,r') &= \frac{1}{2\pi} \int_{0}^{-i \infty}
\left[\frac{f_{+}( \omega e^{2 \pi i},r)}{A(\omega e^{2 \pi i})}
  - \frac{f_{+}( \omega,r)}{A(\omega)}\right]\nonumber \\
  &\quad\quad\quad\quad \times
 f_{-}(\omega,r') p(\omega,r') e^{- i \omega T} d \omega.
\end{align}
This expression, although not particularly helpful in revealing
interesting features of the backscattering, can be evaluated
numerically to obtain an exact result valid for all observers. A more
simplified expression can be obtained if the position of the observer
is assumed to be far away from the horizon. We take the second set of
approximate solutions for~$f_{-}$ constructed from
Eqns.~\eqref{eq:WhittSoln34a}-\eqref{eq:WhittSoln34b}, obtaining
\begin{align} \label{eq:tailsoln}
  f_{-} \approx C_1 Z M_{2 i \omega M,\frac{1}{2} \sqrt{(2 l+1)^2
      -48 \omega ^2 M^2}}(-2 i \omega r),\nonumber\\ f_{+} \approx
  C_2 Z W_{2 i \omega M,\frac{1}{2} \sqrt{(2 l+1)^2 -48 \omega ^2
      M^2}}(-2 i \omega r),
\end{align}
where
\begin{align}
Z = \sqrt{\frac{2 M}{r}} \left(\frac{r}{2 M}
-1\right)^{-\frac{1}{2}+2 i \omega M} e^{-i \omega h(r)}.
\end{align}
The constants~$C_1$ and~$C_2$ can be evaluated from the specific
choice of normalization in the boundary conditions but we do not
evaluate them here since they are absent from the final expression for
the Green's function.

The confluent hypergeometric function of Tricomi or alternatively, the
Whittaker-W function, has a branch cut along the negative
imaginary $\omega$-axis. The~$f_+$ solution is responsible for the
branch cut in the Green's function. The properties of the asymptotic
solutions across the branch cut make them convenient to use. We will
use the general result obtained from Eqn.~(13.14.12)
of~\cite{NIST:DLMF}
\begin{align} \label{eq:WhittakerBC}
W_{a,b} (x e^{2 \pi i}) &= \frac{2 \pi i}{\Gamma(1 + 2 b)
	\Gamma(\frac{1}{2} - b - a)} M_{a,b}(x) \nonumber\\
&\quad- e^{-2 \pi b i} W_{a,b} (x),
\end{align}
to obtain a relation between~$f_+(\omega e^{2 \pi i})$,
$f_+(\omega)$ and~$f_-(\omega)$
\begin{align}
f_{+}(\omega e^{2 \pi i},r) \approx \xi (\omega) f_{+}(\omega,r)
+ \frac{C_2}{C_1} \chi(\omega) f_{-}(\omega,r),
\label{eq:branchrelation}
\end{align}
where
\begin{align} \label{eq:xi}
\xi(\omega) &= -e^{- \pi i \sqrt{(2 l+1)^2-48 \omega ^2 M^2}}, \nonumber\\
\chi(\omega) &= \frac{2  \pi i }{\Gamma \left(\frac{1}{2}-2 i \omega M
	-\frac{1}{2} \sqrt{(2 l+1)^2-48 \omega ^2 M^2}\right)}\times \nonumber \\
&\quad \frac{1}{\Gamma \left( \sqrt{(2 l+1)^2-48 \omega^2 M^2}+1\right)}.
\end{align}
Using Eqn. (\ref{eq:branchrelation}) and noting that the~$f_{-}$
solution does not have a branch cut along the negative
imaginary-$\omega$ axis, we can show that $A(\omega e^{2 \pi i}) =
\xi(\omega) A(\omega)$. This can be used to further simplify the
approximate Green's function
\begin{align} \label{eq:BCequation}
G^{B}(T,r,r') &\approx \frac{1}{2 \pi} \int_{0}^{-i \infty}f_{-}(\omega,r)
f_{-}(\omega,r') \times \nonumber \\
&\quad\quad\quad
\frac{B(\omega)}{A(\omega)}  p(\omega,r')e^{- i \omega T} d \omega.
\end{align}
The standard weighted Wronskian of the~$f_{-}$ and~$f_{+}$
solutions~$A(\omega)$ and~$B(\omega)$ are
\begin{align} \label{eq:tailwronskian}
A(\omega) &= - 2^{3-4 i \omega M} i  \omega C_1 C_2 \mathcal{W}_{M,W}(\omega)
M^{2-4 i \omega M}, \nonumber \\
B(\omega) &= \frac{C_2}{C_1} \frac{\chi (\omega)}{\xi (\omega)},
\end{align}
where the Wronskian between~$M_{a,b}$ and~$W_{a,b}$,
denoted by~$\mathcal{W}_{M,W}$ with respect to the variable~$-2 i
\omega r$ can be written as a ratio of two gamma
functions~\cite{NIST:DLMF}
\begin{align} \label{eq:wronmw}
\mathcal{W}_{M,W}(\omega) = -\frac{\Gamma \left(\sqrt{(2 l+1)^2
-48 \omega ^2 M^2}+1\right)}{\Gamma \left(\frac{1}{2} -2 i \omega M
+\frac{1}{2} \sqrt{(2 l+1)^2-48 \omega ^2 M^2 }\right)}.
\end{align}

\subsubsection*{General and mid frequency}

To calculate the effect of backscattering at arbitrary times for an
asymptotic observer, we can write down a general expression for the
branch cut contribution to the Green's function
\begin{align}
  &G^{B}(T,r,r') \approx - \int_{0}^{-i \infty}
  \frac{e^{-i \left(\omega T + \omega \Xi - \pi  \zeta \right)} \Gamma
    \left(1/2 + \zeta/2 - \sigma \right)
  }{2 \omega  \Gamma \left(1/2 - \zeta/2 - \sigma \right)
    \Gamma \left(\zeta +1\right)^2}\times \nonumber \\
  & \quad \left(\frac{r}{r'}\right)^{-1+\sigma }
  M_{\sigma ,\frac{1}{2}\zeta}(- \sigma r / M )
  M_{\sigma ,\frac{1}{2} \zeta}(-\sigma r' / M ) d\omega, 
\end{align}
where~$\zeta = \sqrt{(2 l+1)^2-48 \omega ^2 M^2}$, $\sigma = 2 i
\omega M$ and~$\Xi(r,r') = h(r) - h(r')$. This expression may be used
as a sanity check for the low and high frequency approximations to the
tail signal.

\subsubsection*{Low frequency}

The late time behavior of the tail is attributed to the low frequency
asymptotics of the approximate Green's function. Hence, in addition to
the approximation for asymptotic observers, we assume~$|\omega M| \ll
1$. This leads to a simplification of the~$f_{-}$ solution and~$\xi$
in Eqns.~\eqref{eq:tailsoln} and~\eqref{eq:xi} respectively
\begin{align}
  f_{-} &\approx \frac{2 M C_1}{r}
  e^{i \omega \Delta(r)}M_{0,l + \frac{1}{2}}(-2 i \omega r), && \xi = 1,
\end{align}
where~$\Delta(r) = 2 M \log (r) - h(r)$. The ratio of two Gamma
functions show up in the expression for the Green's function. This can
be simplified in the low frequency regime yielding
\begin{align}
  \frac{\Gamma (l+1-2 i \omega M)}{\Gamma (-l-2 i \omega M)}
  \approx 2 (-1)^{-l+\frac{3}{2}}  l! \Gamma (l+1) \omega M.
\end{align}
Using these results, we can write down two equivalent expressions,
either as an integral of two Whittaker functions with an exponential
\begin{align}
&G^{B}(T,r,r') \approx -\frac{(-1)^{-l} 2^{-4 l-2} \pi i M r'}
{\Gamma \left(l+\frac{3}{2}\right)^2 r} \int_{0}^{-i \infty}
M_{0,l+\frac{1}{2}}(-2 i \omega r ) \nonumber \\
&\quad\times \ M_{0,l+\frac{1}{2}}(-2 i \omega r')
e^{-i \omega  \big(T-2 h(r')- \Delta(r) - \Delta(r')+\kappa(r')\big)} d \omega, 
\end{align}
or alternatively, as an integration of two Bessel functions with an
exponential
\begin{align}
&G^{B}(T,r,r') \approx \frac{2 i \pi  M r'^{3/2}}{\sqrt{r}}
\int_{0}^{-i \infty} \omega  J_{l+1/2}(\omega r)  J_{l+1/2}(\omega r')
\nonumber \\
&\quad\times e^{-i \omega  \big(T -2 h(r') - \Delta(r) - \Delta(r') + \kappa(r')\big)} d \omega,
\end{align}
with~$\kappa(r') = 4M \log(r')$. Both of these integrals are in their
standard forms and can be evaluated following eqns.~(7.622.3)
and~(6.626.1) of~\cite{GradshteynRyzhik}, so that
\begin{widetext}
\begin{align} \label{eq:BCGFother}
G^{B}(T,r,r') &\approx -\frac{4 (-1)^{l} \sqrt{\pi } M r^l r'^{l+2}
\Gamma (l+2)}{\Gamma \left(l+\frac{3}{2}\right)}
\frac{1}{(T -r_\star - r'_\star +\kappa(r') + \Xi(r,r'))^{2 l + 3}} \\
&\times F_{\Lambda} \left(2 l + 3; l +1, l +1; 2 l + 2, 2 l + 2;
- \frac{2 r}{T - r_\star -r'_\star +\kappa(r') + \Xi(r,r')},
- \frac{2 r'}{T - r_\star -r'_\star +\kappa(r') + \Xi(r,r')}\right), \nonumber 
\end{align}
\begin{align}
G^{B}(T,r,r') &\approx - \sum_{k=0}^{\infty} \frac{(-1)^l 4^{-l-k} \pi 
 M  r'^{l+2} r^{l+2 k} \Gamma (2 l+2 k+3) \, _2F_1\left(-l-k
 -\frac{1}{2},-k;l+\frac{3}{2};\frac{r'^2}{r^2}\right)}
{k! \Gamma \left(l+\frac{3}{2}\right) \Gamma
\left(l+k+\frac{3}{2}\right)(T - 2 h(r') - \Delta(r)
- \Delta(r') +\kappa(r'))^{2l+2k+3}}, \label{eq:BCGF2} 
\end{align}
\end{widetext}
where~$r_\star = r + 2 M \log(r)$, $F_{\Lambda}$ is the hypergeometric
function of two variables and, as before, $_2F_1$ is the Gauss
hypergeometric function. These equations can further be simplified by
using series representations for hypergeometric functions, whose
arguments are suppressed here for brevity,
\begin{align}
_2F_1 = \sum_{s=0}^{\infty} \frac{(-l-k-1/2)_s (-k)_s}{(l+3/2)_s s!}
\left(\frac{r'}{r}\right)^{2s}, \label{eq:2F1}
\end{align}
\begin{align}
&F_{\Lambda} = \sum_{s=0}^{\infty} \sum_{k=0}^{\infty} \frac{(2l+3)_{s+k}
(l+1)_s (l+1)_k}{(2l+2)_{s} (2l+2)_k s! k!} \nonumber \\
&\times \left(\frac{-2 r}{T - r_\star -r'_\star+\kappa + \Xi}\right)^s
\left(\frac{-2 r'}{T - r_\star -r'_\star+\kappa+ \Xi}\right)^k, \label{eq:FLambda}
\end{align}
where Eqn.~\eqref{eq:2F1} is valid when~$|r'/r| < 1$
and Eqn.~\eqref{eq:FLambda} is valid when,
\begin{align}
\left| \frac{-2 r}{T - r_\star -r'_\star+\kappa + \Xi} \right|
+ \left| \frac{-2 r'}{T - r_\star -r'_\star+\kappa + \Xi} \right| < 1. \nonumber
\end{align} 
We note that the condition for validity of Eqn.~\eqref{eq:BCGFother}
is~$T - r_\star - r'_\star + \Xi(r,r') + \kappa(r')> 0$ and for
Eqn.~\eqref{eq:BCGF2} it is~$T - r_\star - r'_\star + \Xi(r,r') +
\kappa(r')> 1$. This must be kept in mind while convolving~$G^B$ with
the initial data. Also, when considering very late times, powers
of~$\left(T-r_\star - r'_\star + \kappa(r') + \Xi(r,r')\right)^{-1}$
and~$\left(T - 2 h(r') -\Delta(r) - \Delta(r') +
\kappa(r')\right)^{-1}$ can be expanded in an inverse power series
of~$T$ about~$T = \infty$. The slowest decaying mode immediately gives
Price's power law~$G^{B} \sim T^{-2 l - 3}$~\cite{PriceTail}.

\subsubsection*{High frequency}

An approximation for the contribution of the tail at very early times
can be computed by considering a high frequency approximation
to~$G_B$. The computations for the high frequency Green's function
become simple when choosing the other pair of asymptotic solutions,
which follows from Eqn.~\eqref{eq:WhittSoln12}, so that
\begin{align} \label{eq:HFsolns}
  f_{-} &\approx C_1 \left(\frac{r}{2 M}\right)^{-1+2 i  \omega M }
  e^{-i \omega h(r)} M_{2 i \omega  M,\frac{1}{2}}(-2 i \omega  r), \nonumber \\
  f_{+} &\approx C_2 \left(\frac{r}{2 M}\right)^{-1+2 i \omega M}
  e^{-i \omega h(r)} W_{2 i \omega  M,\frac{1}{2}}(-2 i \omega  r).
\end{align}
These expressions lead to simplified forms for the
Wronskian~$\mathcal{W}_{M,W}$, $\chi$ and~$\xi$,
\begin{align} \label{eq:HFvals}
  \mathcal{W}_{M,W} = -\frac{1}{\Gamma (1-2 i \omega M)},\,
  \chi = \frac{2 \pi i }{\Gamma (-2 i \omega M)},\, \xi = 1.
\end{align}
Using these expressions, we can also evaluate the ratio of~$B(\omega)$
and $A(\omega)$ at very high frequencies
\begin{align} \label{eq:HFratio}
  \frac{B(\omega)}{A(\omega)} = -\frac{2^{-1+4 i \omega M}
    i \pi M^{-1+4 i \omega M}}{C_1^2},
\end{align}
and also perform a high frequency expansion for~$f_{-}$ as~$r
\rightarrow \infty$
\begin{align} \label{eq:HFfminus}
  f_{-} \approx &-\frac{(-1)^{3/4} C_1 \sqrt{\omega }
  2^{-2 i M \omega } M^{-\frac{1}{2}-4 i \omega M}}{\sqrt{\pi }}\times
  \nonumber \\
  &  r^{-2+4 i \omega M} \left(2 \omega M^2  -i r\right)
  e^{i \omega  (r + 2M - h(r))}.
\end{align}
Using Eqns.~\eqref{eq:HFsolns}-\eqref{eq:HFfminus} in
Eqn.~\eqref{eq:BCequation}, we obtain the final expression for the
time domain Green's function which must now be convolved with the
initial data
\begin{align} \label{eq:GFHFeqn}
  G^{B}(T,r,r') &\approx -\frac{r'}{4 \pi r M^2 \Upsilon(T,r,r')^2 }
  -\frac{r'}{\pi r^2 \Upsilon(T,r,r')^3 } \nonumber \\
  &\quad -\frac{1}{\pi r \Upsilon(T,r,r')^3}
  -\frac{6 M^2}{\pi r^2 \Upsilon(T,r,r')^4 },
\end{align}
where,
\begin{align}
  \Upsilon(T,r,r') &= T + \Xi(r,r') -r-r' - 4M \log r
  \nonumber \\& \quad +
  4M \log M - 4M.
\end{align}
This expression for the Green's function is valid only
when~$\Upsilon(t,r,r') > 0$. Note that the expression in
Eqn.~\eqref{eq:GFHFeqn} is subtle to use in practice because of the
interaction between the validity of the approximation and the domain
of integration, and is hence avoided in comparing with the numerics
later in the paper.

\subsection{Contribution from the high-frequency arc}\label{subsection:directGF}

We now construct the part of the Green's function which comes from the
high-frequency arc, that is when~$|\omega| \rightarrow \infty$. This
gives the part of the signal coming from direct transmission and in
the asymptotic region should reduce to the propagator in flat space.

To derive this result, we write down the Green's function which is
constructed from the Whittaker solutions in Eqn.~\eqref{eq:HFsolns}, 
\begin{align}
G^{HF} \approx \int_{C} \frac{ W_{2 i \omega M,\frac{1}{2}}(-2 i \omega r)
M_{2 i \omega M,\frac{1}{2}}(-2 i \omega r')}{4 \pi i \omega } \nonumber \\
\times\Gamma (1-2 i \omega M)  \left( \frac{r'}{r}\right)^{1-2i\omega M}
e^{-i \omega (T + \Xi)} d \omega,
\end{align}
for $r' < r$. The other case can be derived in a straightforward
manner. Here~$C$ is the contour over which the integration is
performed. In the asymptotic limit~$r \rightarrow \infty$, the
Whittaker functions can be further simplified as
\begin{align}
M_{2 i \omega M,\frac{1}{2}} &\approx \frac{2^{-2 i \omega M} (-i \omega )^{-2 i \omega M}
e^{-i \omega r' } r'^{-2 i \omega M}}{\Gamma (1-2 i \omega M)} \nonumber \\
&\quad-\frac{2^{2 i \omega M} (i \omega )^{2 i \omega M} e^{i \omega r'} r'^{2 i \omega M}}
{\Gamma (1 + 2 i \omega M)}, \nonumber\\
W_{2 i \omega M,\frac{1}{2}} &\approx 2^{2 i \omega M} (-i \omega )^{2 i \omega M}
e^{i \omega r} r^{2 i \omega M}.
\end{align}
In the high-frequency limit, Stirling's formula can be employed for
the Gamma function~\cite{NIST:DLMF},
\begin{align}
\Gamma(a \omega + b) \approx \sqrt{2 \pi} e^{-a \omega}
(a \omega)^{a \omega + b - 1/2}, && |\arg(\omega)| < \pi.
\end{align}
which is valid for $a > 0$ and $b \in \mathbb{C}$. The high-frequency
asymptotic Green's function can be written as
\begin{align}
&G^{H}(T,r,r') \approx \frac{r'}{4 \pi  i r} \int_{C}
\frac{e^{-i \omega  (T + \Xi -r+r'-4M \log r + 4M \log r')}}{ \omega }
d\omega \nonumber\\
& + \frac{r'}{4 \pi r} \int_{C} \frac{ e^{-i \omega
(T + \Xi -r-r'-4 M \log r - 4 M + 4 M \log M)}}{\omega} d \omega.
\end{align}
The choice of contour~$C$ is motivated by the discussion
in~\cite{Andersson1997}. As we have assumed~$r$ to be very large, we
see that only the first term contributes when~$r - r' - 4M \log r \leq
T + \Xi \leq r + r' + 4M \log r'$. Taking a contour~$C$ in the upper
half of the~$\omega$ plane, the leading order term in the Green's
function can be written down in terms of the Heaviside function
\begin{align}
&G^{H} (T,r,r') \approx -\frac{r'}{2 r}
\mathbb{H}(T + \Xi -r+r'-4M \log r + 4M \log r').
\end{align}
For the case of Kerr-Schild coordinates, when convolving with the
initial data, the lower limit of the integration~$r'_{l}$ is obtained
by solving for~$r'$ in
\begin{align}
r' + 4 M \log r' \approx r + 4 M \log r  - t.
\end{align} 
The scalar field response from the initial data as seen by an observer
at fixed~$r$ can then be calculated as
\begin{align}
  K_{l,m}(t,r) = -\int_{r'_{l}}^{r+t} \frac{r'}{2 r} \partial_t K_{l,m}(0,r') dr'
- \frac{r'_{l}}{2 r} K_{l,m}(0,r'_{l}).
\end{align}

\section{Numerical results}\label{Section:Numerics}

In the second part of the paper, we numerically evolve a massless
scalar field and compare with the results obtained from the first part
of the paper. After a brief overview of our pseudospectral NR
code~\verb|bamps| and the~\verb|scalarfield| project in
sections~\ref{subsection:Numerical_Setup}
and~\ref{subsection:Numerical_ID}, we record the main results of our
paper in two separate sections for the QNM and tails.

\subsection{Numerical setup}\label{subsection:Numerical_Setup}

The~\verb|bamps| code~\cite{bamps1,bamps2,bamps3,bamps4,bamps5,bamps6}
is a massively parallel multipatch pseudospectral code for numerical
relativity. The code is written in~C with specific algebra-heavy
components generated by Mathematica scripts. In the present work we
use this tool to solve the wave equation in a fixed Schwarzschild
background. Since \verb|bamps| is primarily designed to treat first
order symmetric hyperbolic systems we therefore start by reducing to
first order as
\begin{align}
\p_t\Phi&=\alpha \Pi+\beta^i\chi_i,\nonumber\\
\p_t\chi_i&=D_i(\alpha\Pi)+\alpha\gamma c_i+\Lie_\beta\chi_i,\nonumber\\
\p_t\Pi&= D^i(\alpha\chi_i)+\alpha K \Pi
+\gamma\beta^ic_i+\Lie_\beta\Pi,\label{eq:WE_11}
\end{align}
subject to the spatial reduction constraint
\begin{align}
c_i\equiv\p_i\Phi-\chi_i=0.\label{eq:WE_11_constraint}
\end{align}
The purpose of the parameter~$\gamma\geq0$ is to damp inevitable
violations of this constraint. The~\verb|scalarfield| project is
coupled to our metric evolution scheme and has been tested on each of
our domains, but in the present context, as we excise the black hole
region, we work exclusively with nested cubed-shell grids. In this
section we employ the standard~$3+1$ notation~\cite{alcubierre,BS} for
the future pointing unit normal vector, lapse, shift, spatial
covariant derivative and extrinsic curvature. The values for these
quantities can be read off from the background metric. The independent
non-trivial values are
\begin{align}
  \alpha&=\frac{1}{\sqrt{1+2M/r}},\quad &
  \beta^r&=\frac{2M/r}{1+2M/r}  ,\nonumber\\
  \gamma_{rr}&=1+2M/r,\quad &
  \gamma_{\theta\theta}&=r^2,\nonumber\\
  K_{rr}&=\frac{-2M(M+r)}{\sqrt{r^5(2M+r)}},\quad &
  K_{\theta\theta}&=2M\sqrt{\frac{r}{2M+r}}\,,
\end{align}
in spherical polars. In the code these are transformed to our global
Cartesian basis in the obvious manner. When evolving the system
coupled to GR, we use first order reduction variables in place of
taking derivatives of metric components so that the scalarfield and
gravitational field equations remain minimally coupled from the PDEs
point of view. The characteristic variables for the system are
\begin{align}
  u^{\pm}&=\pm s^i\chi_i+\Pi+\gamma \Phi,
  \quad u^\beta_i=(\delta^j{}_i-s^js_i)\chi_j,\nonumber\\
  u^0&=\Phi,
\end{align}
with geometric speeds~$-\beta^is_i\mp\alpha$,~$-\beta^is_i$ and~$0$
respectively, where~$s^i$ denotes an arbitrary unit spatial
vector. The computational domain is divided into subpatches, each of
which is discretized in space using a Gauss-Lobatto grid with a
Chebyshev basis. Thus spatial derivatives are ultimately approximated
with matrix multiplication. The equations of motion~\eqref{eq:WE_11}
are then integrated in time using a standard fourth order Runge-Kutta
method. Data is communicated into a given patch from its neighbors by
weakly imposing equality of incoming characteristic fields using a
penalty method. At the outer boundary we impose
\begin{align}
  r^{-2}L^\mu\p_\mu\big(r^2(u^+-\gamma \Phi)\big)&=0,\nonumber\\
  (\delta^j{}_i-s^js_i)s^k\p_{[j}c_{k]}&=0,
\end{align}
with~$s^i$ here the spatial outward pointing unit normal to the
domain, and~$L^\mu=n^\mu+s^\mu$ is an outward pointing
null-vector. These conditions are constraint preserving and control
incoming radiation. It should be noted, however, that we typically
ensure that the outer boundary is causally disconnected from the
region of spacetime we study, so that at the continuum level we are
effectively considering the IVP rather than the IBVP. Presently we
work entirely in spherical symmetry, so we use the cartoon
method~\cite{cartoon1,cartoon2} to reduce the number of spatial
dimensions to one. Apart from the fact that this allows us to rapidly
produce many data sets on a large desktop machine, enforcing explicit
spherical symmetry ensures that only the $l=0$ mode is excited as our
study primarily involves the effect of overtones on the signal. Our
code is MPI parallel; large jobs are run on a multi-core
workstation. The results of these simulations are compared with our
standalone Green's function code written in Python. More details
of~\verb|bamps| can be found in~\cite{bamps2}.

\subsection{Initial data}\label{subsection:Numerical_ID}

For the purposes of this paper, we consider only initial data which is
spherically symmetric. This is not a restriction in itself, since the
analysis followed can be extended to non-spherical initial data in a
straightforward manner. For initial data, we provide the value of the
scalar field~$\Phi$ at~$t = 0$ and
\begin{align}
\Pi = \alpha^{-1} (\partial_t \Phi - \beta^r \partial_r \Phi),
\end{align}
also at~$t = 0$. Here~$\alpha$ and~$\beta^i$ are the lapse and shift
respectively. The four different types of initial data used in our
runs are listed below;

\begin{description}
\item[Gaussian profile~I (type A).] This is the simplest type of
  initial data with the following profile
  \begin{align}
    \Phi &= A e^{-(r-r_0)^2/\sigma^2}, \qquad \Pi = 0.
  \end{align}
  Here~$A$ is the amplitude of the scalar field, $r_0$ is the position
  of the peak of the Gaussian and~$\sigma/\sqrt{2}$ is the standard
  deviation.
\item[Gaussian profile~II (type B).] The second type of initial data is
  the purely ingoing Gaussian pulse in the Minkowski spacetime
  \begin{align}
  \Phi &= \frac{A}{r} e^{-(r-r_0)^2/\sigma^2}, \nonumber \\ \Pi &= -
  \frac{2 A}{r \sigma^2} (r - r_0) e^{-(r-r_0)^2/\sigma^2}.
  \end{align}
Here $A/r$ is the amplitude of the scalar field, $r_0$ is the position
of the peak of the Gaussian and $\sigma/\sqrt{2}$ is the standard
deviation. On a Schwarzschild background, this data is `mostly
ingoing'.
\item[Sine Gaussian profile (type C).] The scalar field profile is
  given by
  \begin{align}
  \Phi &= \frac{A}{r} e^{-(r-r_0)^2/\sigma^2} \sin (\omega r + \phi_0),&\quad
  \Pi = 0.
  \end{align}
Here,~$A/r$ is the amplitude of the scalar field and~$r_0$ is the peak
of the scalar field while~$\omega$ and~$\phi_0$ are the frequency and
phase of the oscillating frequency. 

\item[Pure QNM initial data profile.] We would also like to evolve pure
  QNM data of the form
  \begin{align}
  \Phi &= A e^{-i \omega r} H_C\left(\Theta,1-\frac{r}{2M}\right),\nonumber\\
  \Pi &= \frac{(\beta^r - 1)}{\alpha} i \omega A e^{-i \omega r}
  H_C\left(\Theta,1-\frac{r}{2M}\right) \nonumber \\
  &\quad- A e^{- i \omega r} \frac{\beta^r}{\alpha} \frac{d}{d r}H_C
  \left(\Theta,1-\frac{r}{2M}\right),
  \end{align}
with~$\Theta = \{4i \bar{\omega},-4i \bar{\omega},0,-8
\bar{\omega}^2,8 \bar{\omega}^2\}$ being the parameters of the CHE,
the construction of which is detailed in
section~\ref{subsection:QNMs}. The field profile of this type of data
is bounded at the horizon but not at spatial infinity. To ensure that
the field remains smooth during numerical evolution, data at the outer
boundary is initially kept to be zero, at least to machine precision
by multiplying both~$\Phi$ and~$\Pi$ with a smooth cutoff
function. More details are provided in
section~\ref{subsub:qnmnumerics}.
\end{description}

\begin{figure*}[!th] 
\centering
\includegraphics[width=\textwidth]{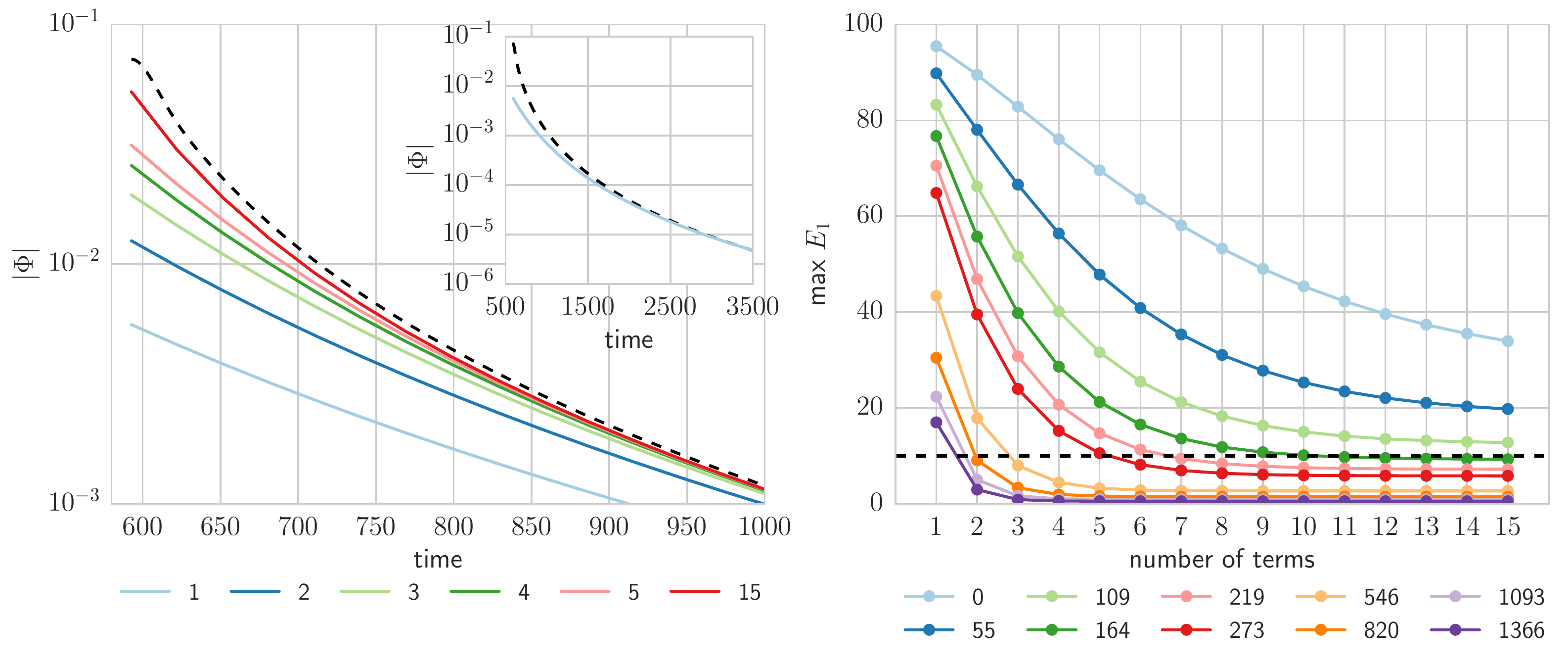}
\includegraphics[width=\textwidth]{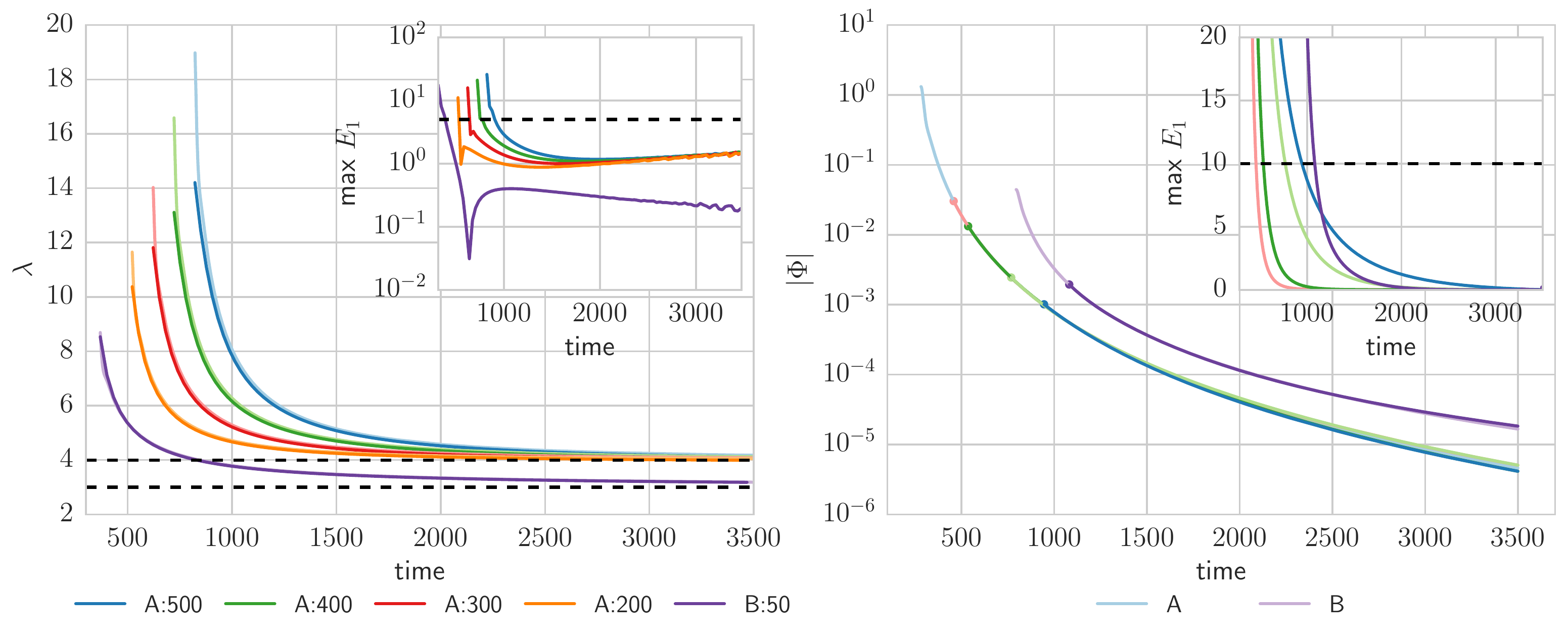}
\caption{\textit{Top row, left}: Comparison between the scalar field
  time-series extracted at~$r \simeq 500$ (in dotted line) with those
  obtained from the approximate Green's function for a simulation with
  Gaussian I type initial data. The colors indicate the different
  number of terms considered in the summation in
  Eqn.~\eqref{eq:BCGF2}. The inset figure shows that at late times,
  the slowest decaying~$t^{-4}$ term dominates. \textit{Top row,
    right}: Variation of the maximum~$E_1$ between~$\Phi_\textrm{b}$
  and~$\Phi_\textrm{gf}$ at different starting times, mentioned in
  units of M for Gaussian II type data. \textit{Bottom row, left}: A
  comparison between the LPI for two simulations at different observer
  positions mentioned in the legend, in units of~$M$. The inset plots
  the maximum~$E_1$ for the corresponding LPI
  comparison. \textit{Bottom row, right}: Results of the tail fits to
  two different types of initial data when considering~$q=1,2,3,4$ and $q = 4$ in
  Eqn.~\eqref{eq:tailmodel} respectively. The inset plot shows the variation of the
  maximum $E_1$ with time for the different fits with the dotted line
  denoting $10 \%$. The fits get progressively better as~$q$
  increases. In both plots, `A' corresponds to a simulation with
  Gaussian I data and `B' to Gaussian II type data.}
\label{fig:tail1}
\end{figure*}


\begin{figure*}[!th]
\centering
\includegraphics[width=\textwidth]{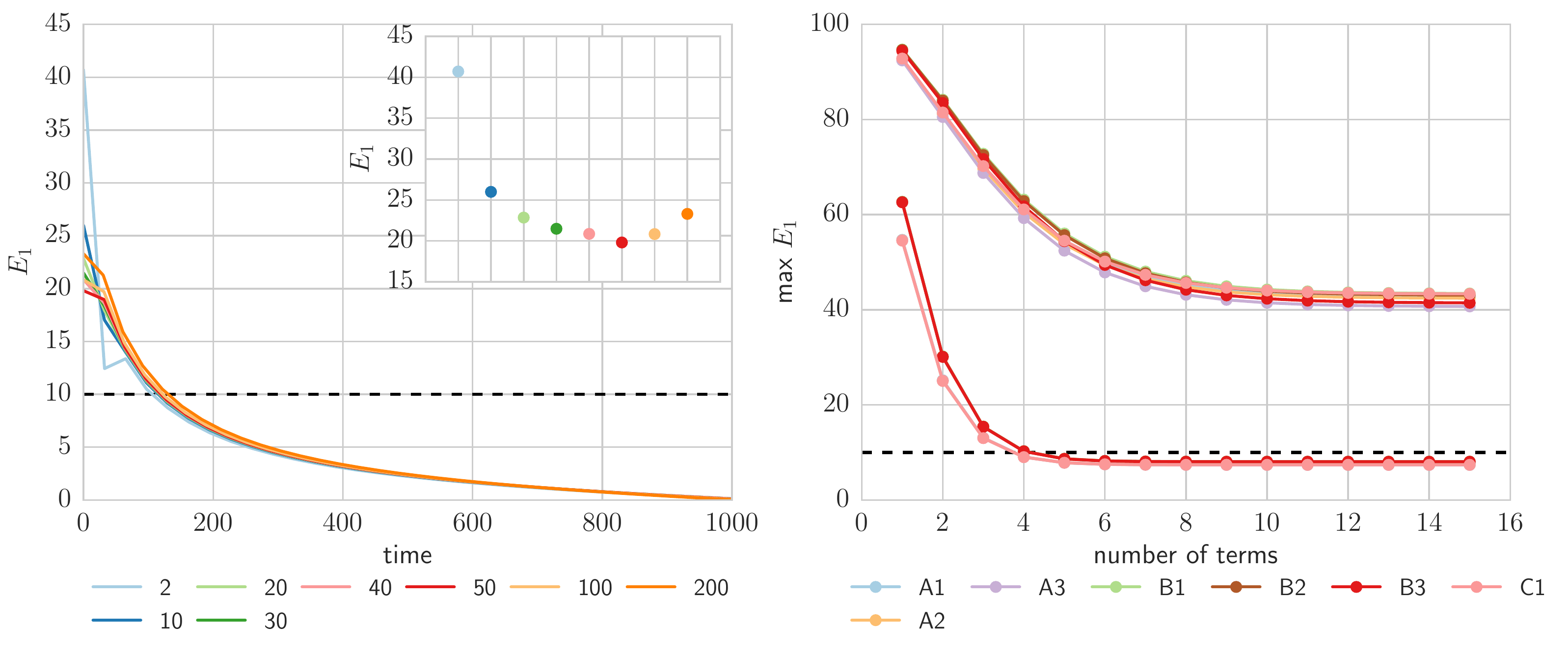}
\caption{\textit{Left}: Variation of the pointwise percentage
  error~$E_1$ between the numerical signal and the late time Green's
  function result as a function of time for scalar field extracted at
  radii~$r \simeq$ $2$, $10$, $20$, $30$, $40$, $50$, $100$ and $200
  M$ for Gaussian I data. The inset plot shows the values of $E_1$ at
  the beginning of the tail for the same time-series. \textit{Right}:
  Variation of the maximum~$E_1$ with respect to time for an observer
  at~$r \simeq 2 M$ with different number of terms in the Green's
  function in Eqn.~\eqref{eq:BCGF2} at two starting times for all~$7$
  simulations. At intermediate starting times, the maximum error can
  be kept to less than~$10 \%$ if~$5$ or more terms are considered in
  the Green's function. The legend shows the three type A, three type
  B and one type C simulation used in the analysis.}
\label{fig:tail2}
\end{figure*}

\subsection{Tests on tails}\label{subsection:tailnumerics}

We first test the expressions for late time tails using a set of~$7$
high-resolution simulations with Gaussian I, II and sine-Gaussian type
initial data. For each simulation, the scalar field information is
extracted for~$10$ different observers outside the event horizon whose
positions are approximately at~$r \simeq
2$,~$10$,~$20$,~$30$,~$40$,~$50$,~$100$,~$200$,~$300$,~$400$ and ~$500
M$. As mentioned earlier, care must be taken to place the outer
boundary at a sufficiently large radius compared to the position of
the observer and the initial `pulse' to ensure that boundary effects
do not contaminate the time-series in the region of interest. This
problem could be completely avoided by evolving the scalar field in
hyperboloidal coordinates. In the following analysis, what we call the
`tail signal' starts when the QNM ringing ceases to dominate the
signal, and for operational purposes, this starts from the last
extremum of the time-series onwards. This signal is then compared with
our standalone Green's function code which computes the low frequency
contribution of the branch cut using Eqn.~\eqref{eq:BCGF2}, but with a
truncated sum. To quantify the disagreement between the numerical data
and the Green's function result, we define a measure of error
\begin{align}
  E_1(t,r_\textrm{o}) = \left|\frac{\Phi_\textrm{b} -
  \Phi_{*}}{\Phi_\textrm{b}}\right| \times 100,
\end{align}
which is the percentage error at time~$t$ as seen by an observer at
$r_\textrm{o}$. Here~$\Phi_\textrm{b}$ is the numerical signal which
has~$N$ points while~$\Phi_*$ is $\Phi_\textrm{gf}$
or~$\Phi_\textrm{f}$, the signal computed either from the approximate
Green's function or derived from fitting a model respectively.

One of our aims in this section is to construct a model for the tail
signal as a superposition of power laws. To find out the number of
terms needed to faithfully represent the numerical result, we generate
the first~$15$ terms of the approximate Green's function as in
Eqn.~\eqref{eq:BCGF2} and compute the maximum percentage error
between~$\Phi_\textrm{b}$ and~$\Phi_\textrm{gf}$ while cumulatively
adding more terms in~$\Phi_\textrm{gf}$. Additionally, the starting
time for computing the mismatch is varied to compute the maximum~$E_1$
across early, intermediate and late time tails separately.

\begin{figure*}[!th] 
\centering
\includegraphics[width=\textwidth]{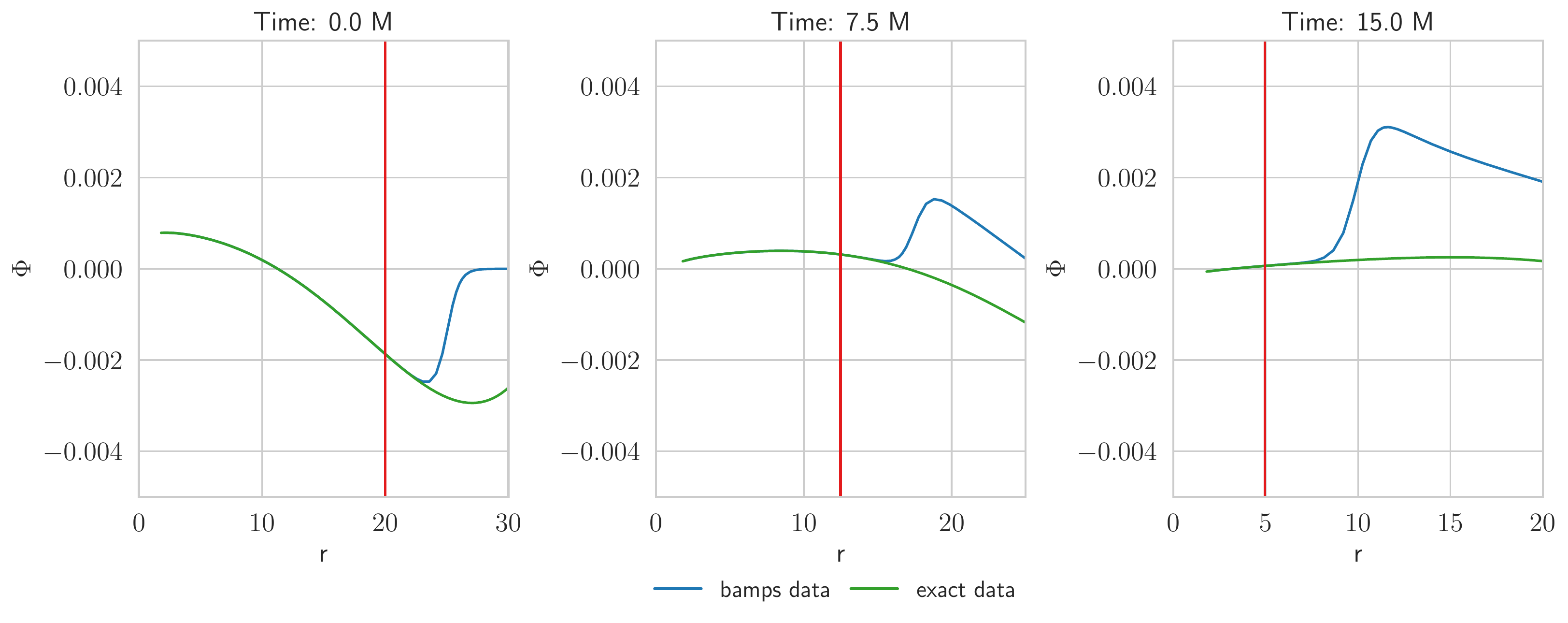}
\includegraphics[width=\textwidth]{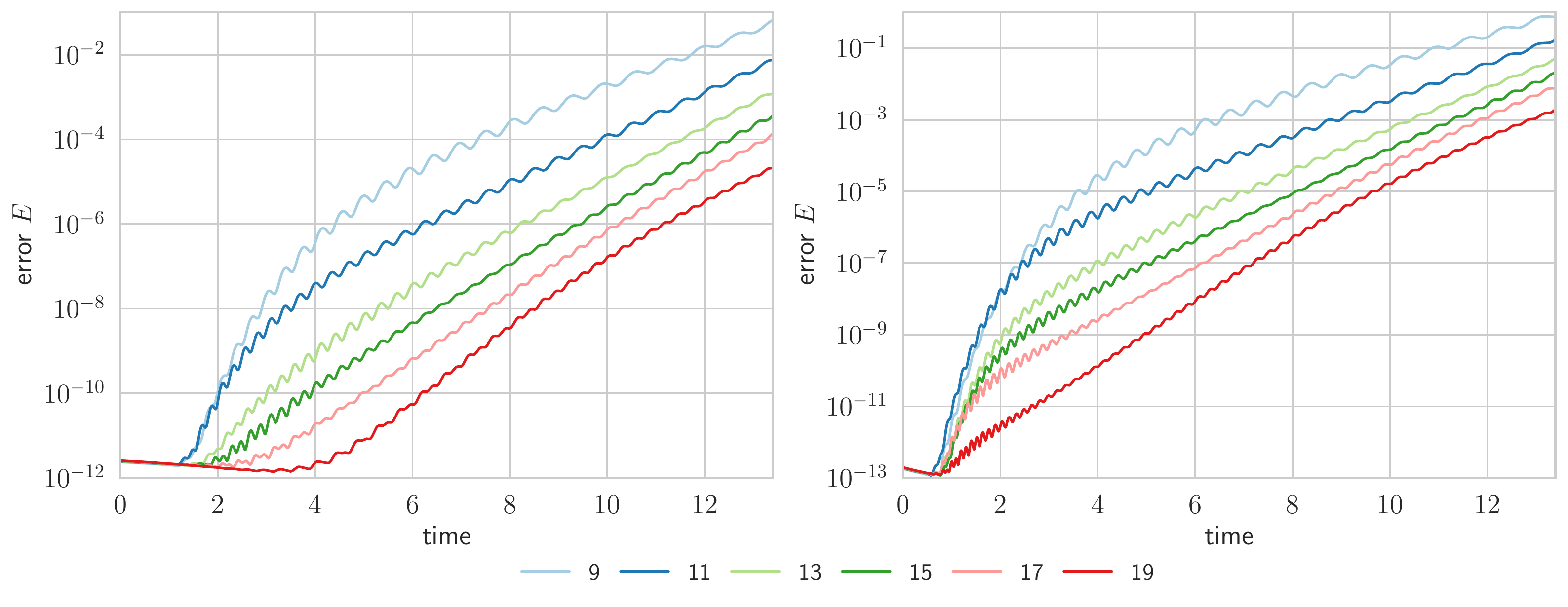}
\caption{\textit{Top row, from left to right}: A comparison between
  the pure QNM solution and the numerical data for the~$n = 0$ mode at
  three different times. The left of the vertical red line denotes the
  region up to which the numerical data and the pure QNM data must
  agree. \textit{Bottom row, from left to right}: Convergence plot
  for~$\Phi$ and~$\Pi$ respectively for data prepared from the $n=1$
  mode with the error computed from
  Eqn.~\eqref{eq:convergenceerror}. The colors represent the different
  resolutions of the simulations.}
\label{fig:exactID}
\end{figure*}

At very late times, we see that the maximum~$E_1$ does not vary
significantly with the addition of a few terms irrespective of where
the observer is located and both can be kept to less than~$10
\%$. This however becomes progressively worse at earlier times and may
be improved by adding more terms in our approximation of the Green's
function as can be seen in the top row, left of
Fig.~\ref{fig:tail1}. For the simulations considered, the intermediate
tail onwards can be described to an accuracy of~$< 10 \%$ error if at
least the first~$5$ terms are considered in
Eqn.~\eqref{eq:BCGF2}. This is demonstrated in an example simulation
in the top row, right of Fig.~\ref{fig:tail1} and is the rule of thumb
followed when building our model.

At very late times, we observe only the effects of the~$m=0$ term
within Eqn.~\eqref{eq:BCGF2} on the signal. This is reflected in the
local power index (LPI)~$\lambda$ of the time-series defined
as~\cite{Harms_2013}
\begin{align}
|\Phi| = A t^{-\lambda}, &\quad
\lambda = - \frac{\partial \log |\Phi|}{\partial \log t}.
\end{align}
A comparison between the LPI computed from~$\Phi_\textrm{b}$
and~$\Phi_\textrm{gf}$ again shows that they are in good agreement for
intermediate times.  Generally irrespective of the position of the
observer the maximum~$E_1$ between the numerical and analytically
computed~$\lambda$ can be kept smaller than~$10\%$, and the mismatch
is only large at very early times, as can be seen in bottom row, left
of Fig.~\ref{fig:tail1}. For late times,~$\lambda$ approaches~$4$ for
Gaussian I and sine-Gaussian data and~$3$ for Gaussian II data, which
is consistent with Price's law~\cite{PriceTail}. We must note here
that the time derivative of the scalar field is only approximately
zero at large radii for Gaussian I and sine-Gaussian data,
so~$\lambda$ must eventually approach~$3$ if the simulations are
evolved for a much longer time.

We now proceed to fit a sum of tail laws to the data with constant
coefficients
\begin{align} \label{eq:tailmodel}
|\Phi| = \sum_{k=0}^{q} A_{k} t^{-(3+k)}.
\end{align}
While this model is a simplification over the time dependence in
Eqn.~\eqref{eq:BCGF2}, it should work well for late times. A
non-linear least squares fit is performed using the
Levenberg-Marquardt algorithm as implemented in the Python package
\verb|lmfit|~\cite{lmfit} at different starting times to obtain the
coefficients~$A_k$. Non-linear fitting algorithms are sensitive to
initial conditions and can perform poorly if the~$A_k$'s are
initialized with random values. To initialize the first non-zero
coefficient, we make use of the fact that at late times, the slowest
decay dominates the signal and is either the $t^{-3}$ term or the
$t^{-4}$ term depending on the initial data. While obtaining the first
non-zero coefficient is straightforward, it is less clear how to
obtain good guesses for the others. Noting that the other tail
components contribute significantly at earlier times, we initialize
all coefficients with values of the first non-zero coefficient. This
empirical approach works remarkably well in practice.

For each time-series, the starting time is shifted over the entire
signal and the maximum percentage error is recorded for each
position. The maximum~$E_1$ with respect to time between the fit and
the numerical data displays a monotonically decreasing behavior with
starting time. The earliest recorded start time at which the
maximum~$E_1$ is less than or equal to~$10 \%$ is considered the
optimal starting time for the fit. The maximum number of terms~$q$ in
Eqn.~\eqref{eq:tailmodel} is also allowed to vary from~$1$
to~$11$. More of the signal can be modeled with increasing~$q$
until~$q$ is equal to~$5$. The maximum~$E_1$ gets worse as~$q$ is
increased further, and hence the very early part of the signal is not
well represented as a linear combination of different power law tails
with constant coefficients. A representative fit with different number
of terms in the tail model is shown at the bottom row, right of
Fig.~\ref{fig:tail1}.

Finally, we look at how well the asymptotic expressions perform for
non-asymptotic observers. To do this, we compute the pointwise
percentage error between~$\Phi_\textrm{b}$ and~$\Phi_\textrm{gf}$ for
several positions outside the event horizon and find that the
beginning of the tail signal always has a considerable error~($> 10
\%$) irrespective of the simulation and the position of the
observer. However, at intermediate times, the percentage error falls
beneath~$10 \%$ and therefore we propose that the asymptotic tail
expressions may also be used for observers very close to the event
horizon at intermediate and later times. Furthermore, we also look at
the maximum percentage error for signals at~$r \simeq 2 M$ across
the~$7$ simulations and find that it cannot be kept to under~$10 \%$
if the entire tail signal is chosen for the analysis. It is only from
intermediate time onwards that the error~$E_1$ can be kept under~$10
\%$. Fig.~\ref{fig:tail2} left shows the variation of $E_1$
between~$\Phi_\textrm{b}$ and~$\Phi_\textrm{gf}$ with time while
Fig.~\ref{fig:tail2} right shows the maximum percentage error when the
fit is performed from the beginning of the tail signal and from
intermediate times, for different number of terms in the tail model.

\subsection{Tests on QNMs} \label{subsub:qnmnumerics}

\subsubsection{Results from exact solutions} 

Looking at the analytically continued solutions~$f_{-}$ and~$f_{+}$,
we see that at QNM frequencies, these solutions are unbounded at
spatial infinity. This fact makes it difficult to evolve pure QNM type
initial data in our numerical code unless the outer boundary can be
treated appropriately. One suggestion is to have time dependent
boundary conditions at the outer boundary which can be analytically
determined. The alternative is to make the initial data near the outer
boundary of the order of machine precision or less by employing a
smooth cutoff function, \begin{align} C(r,a,r_0) = \frac{1}{2}
    \left(1 - \tanh(a(r-r_0))\right).
\end{align}
This cutoff changes smoothly from~$1$ to~$0$ around~$r_0$, with~$C$
being~$0.5$ at~$r_0$. The steepness of this change is controlled by
the value of~$a$.

The second route is easier to implement and is the one followed
here. Our objective in these experiments is two-fold. First we wish to
obtain an arbitrarily long `ringing time' for observers close to the
event horizon and second to have a ringdown at a specific QNM
frequency. This data can be used to obtain an arbitrarily long ringing
time of a single QNM, or a superposition of QNMs near the event
horizon. Since the ingoing light speed is exactly~$1$ in Kerr-Schild
coordinates, to obtain a ringing duration of~$\Delta t_\textrm{QNM}
\simeq 20 M$, the pure QNM solution and the initial data must match up
until at least~$r \simeq 22 M$. This is achieved in our case by
choosing~$r_0 = 25 M$ and~$a =1$ for the~$n = 0$ mode and~$r_0 = 25 M$
and~$a = 2$ for the~$n = 1$ mode.

A good way to test the correctness of the \verb|scalarfield|
implementation is to perform a convergence test with the initial data
built from the exact solution in the region unaffected by the cutoff
function. The basic steps for implementing such a test are given
below,

\begin{enumerate}[1.]
\item Generate and evolve the modified QNM data on a Schwarzschild
  background at~$5$ different resolutions or more. For our test, we
  choose subpatches with~$9$ to~$19$ points, increasing the number of
  points by~$2$ each time.
\item Compute the pure QNM initial data at a much higher resolution
  than the highest resolution used for the numerical runs. This
  ensures that interpolation errors, which can be problematic, do not
  dominate in the test. We constructed the QNM data from~$r \in
  [1.8,50]$ with~$5\cdot 10^4$ points or more.
\item Interpolate the exact solution on each \verb|bamps| subpatch and
  output at the same times as in our numerical simulations. A
  comparison of the analytically evolved initial data and the
  numerical data for the~$n = 0$ mode at three different times is
  shown in the top row of Fig.~\ref{fig:exactID}.
\item Compute the error~$E$ between the numerical
  result~$\Phi_\textrm{b}$ and the analytic results~$\Phi_\textrm{e}$
  for the first~$P$ subpatches where the data is not affected by the
  cutoff function
  \begin{align} \label{eq:convergenceerror}
  E(t) = \sum_{i = 1}^{P}\int_{i} (\Phi_\textrm{e}(t,r) - \Phi_\textrm{b}(t,r))^2 dr.
  \end{align}
  Here the data on each grid is specified at the Gauss-Lobatto points
  and the weights for the integration on each grid with~$N$ points
  must be calculated from the Chebshev Gauss-Lobatto numerical
  quadrature
  \begin{align}
  w_i = \begin{cases} \sqrt{1 - x^2_i} ({\pi}/{2 N}), \ i = 0,N, \\
  \sqrt{1 - x^2_i} ({\pi}/{N}), \ \text{elsewhere}, \\
  \end{cases}
  \end{align}
  where $x_i$ for each grid are given by,
  \begin{align} \label{eq:GL}
  x_i = - \cos \left(\frac{\pi i}{N - 1}\right).
  \end{align}
\item Plot the error~$E$ as a function of time for each
  resolution. For the test to be successful,~$E$ should decrease with
  increasing resolution. A convergence test for the~$n=1$ mode is
  shown in the bottom row of Fig.~\ref{fig:exactID}.
\end{enumerate}

A comparison between the numerical and analytical solutions at
different times show excellent agreement in the region unaffected by
the cutoff. We use the matrix pencil~\cite{Hua} and Prony
methods~\cite{BertiDA} to fit damped exponentials to the time series
data on the horizon. Since the signal is real, we fit two damped
exponentials~$A e^{i \omega t + i \phi_0}$ for each mode, $\omega$
being the complex QNM frequency. The parameters of the fit provide
very accurate numbers for the QNM frequency, namely~$0.11043074 -
0.10485913 i$~(with less than~$0.01 \%$ error) for the~$n=0$ mode
and~$0.0857 - 0.3472 i$~(with less than~$0.1 \%$~error) for the~$n=1$
mode.

\begin{figure*}[!th] 
\centering
\includegraphics[width=\textwidth]{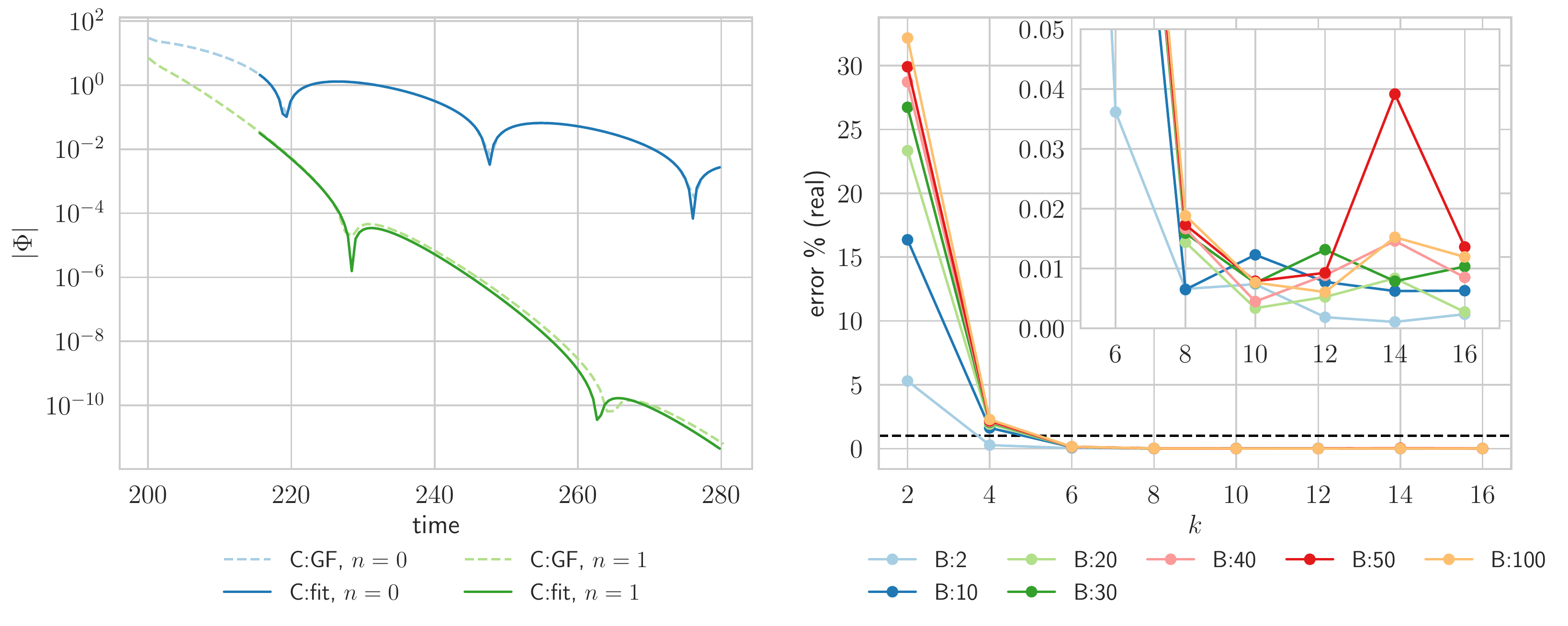}
\includegraphics[width=\textwidth]{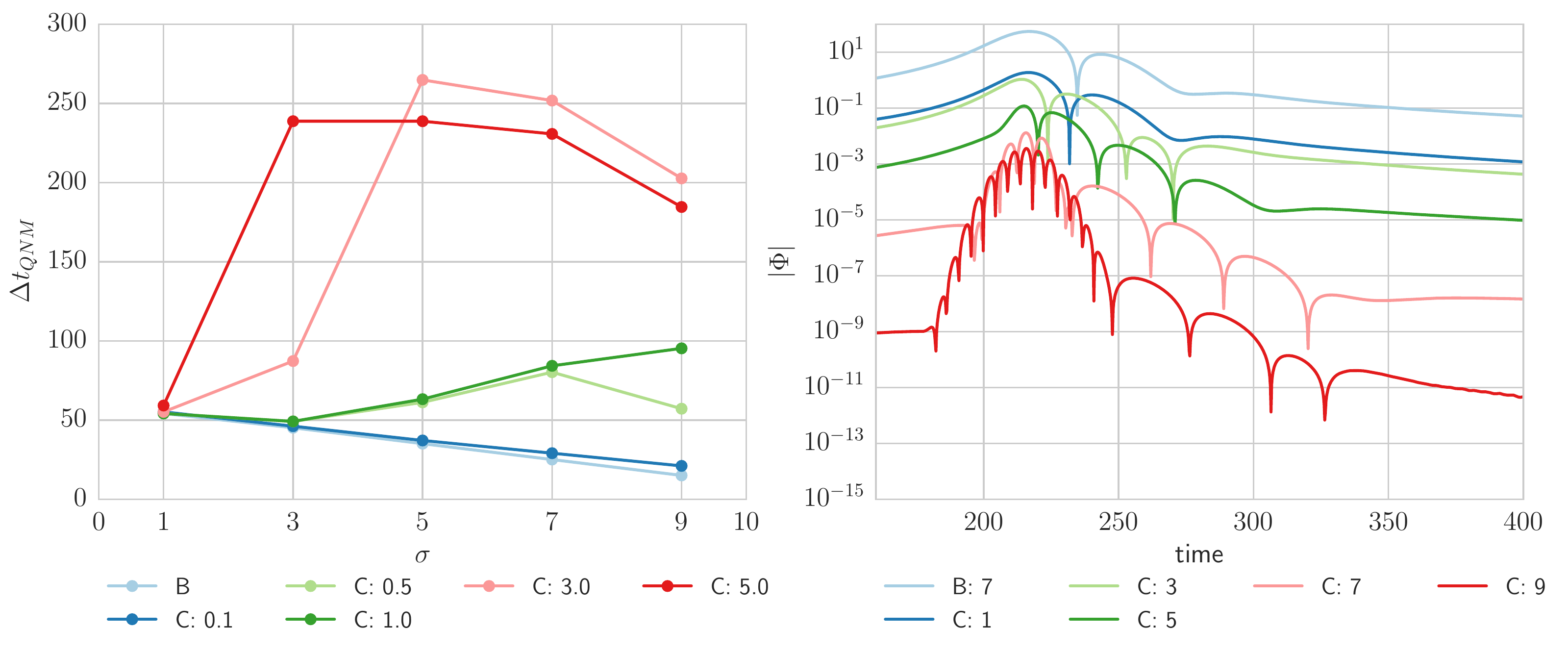}
\caption{\textit{Top row, left}: A comparison between the numerical
  data and the Green's function result for the $n = 0$ and $n = 1$
  mode for sine-Gaussian initial data. \textit{Top row, right}: The
  percentage error in extracting the real part of the~$n = 0$ QNM
  frequency from the fit when different number of terms are considered
  in the fit model in Eqn.~\eqref{eq:qnmmodel}. The legend specifies
  the positions of the observer considered, in units of~$M$ for a
  Gaussian II type simulation. The inset plot zooms in a portion of
  the plot. \textit{Bottom row, left}: Approximate ringing time for
  observers at~$r \simeq 100 M$ calculated from the Green's function
  for Gaussian II and sine-Gaussian initial data for different values
  of~$\sigma$. The legend specifies the values of~$\omega$
  used. \textit{Bottom row, right}: Variation of QNM ringing from
  numerical simulations as seen by an observer at~$r \simeq 100 M$ for
  Gaussian II and sine-Gaussian initial data~($\omega = 1$). The
  legend specifies the values of~$\sigma$ used.}
\label{fig:qnm1}
\end{figure*}

\subsubsection{Results from generic data}

We now test our expressions for the QNM part of the Green's function
using non-specialized initial data. For this, we use the simulations
in section~\ref{subsection:tailnumerics} taking~$6$ observers outside
the black hole at at~$r \simeq 2$, $10$, $20$, $30$, $40$, $50$
and~$100 M$.

The time-series at each of these points must be cropped to include
just the `ringing' part of the signal. To do this, we restrict the
signal to the interval between the first extrema during ringing and
the start of the `tail signal'. During the data analysis, the starting
time for the fit is varied over the signal and the time at which the
normalized modulus square of the difference between the fit and the
numerical data is found to be minimized is chosen as the optimum
starting time for the fit.

The model for the fit is chosen to be a linear combination of~$k$
damped exponentials
\begin{align} \label{eq:qnmmodel}
\Phi_\textrm{mp} = \sum_{j = 0}^{k} A_j e^{(\alpha_j + i \omega_j)t + i \phi_j},
\end{align}
where the fit is performed for the
parameters~$\{A_j,\alpha_j,\omega_j,\phi_j\}$ using the matrix pencil
method~\cite{Hua}. For real signals,~$k$ is chosen to be an even
number and the pencil parameter is kept at one-third the number of
points in the time-series rounded off to an integer value. To ensure
that the frequencies extracted from the signal are reliable, we check
to ensure that the values of~$|\omega_j| - i |\alpha_j|$ which must
occur in pairs, do not differ from each other by more than~$10^{-4}$
in both the real and imaginary part. Since the algorithm is designed
for complex signals in general,~$\Phi_\textrm{mp}$ may have an
imaginary component and this ensures that it is kept small. A summary
of the steps to implement this algorithm is provided
in~\cite{BertiDA}.

\begin{figure*}[!th] 
\centering
\includegraphics[width=\textwidth]{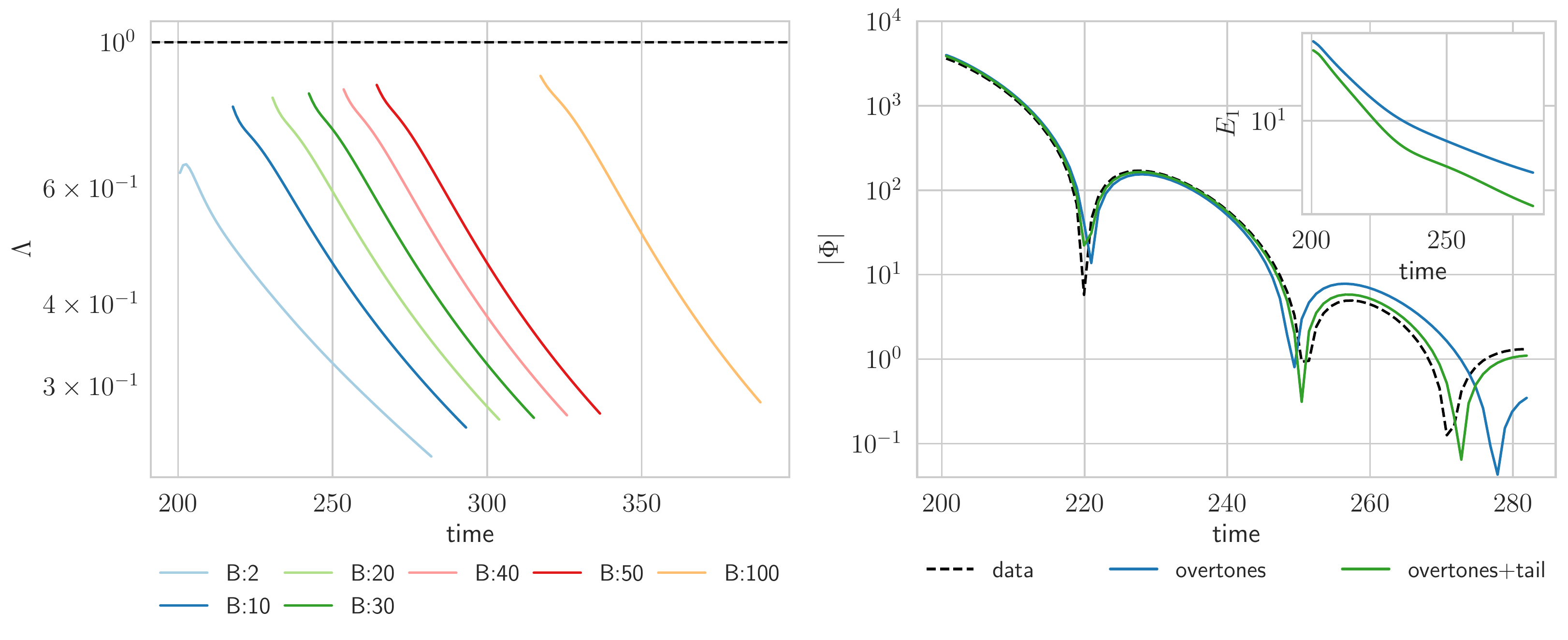}
\caption{\textit{Left}: The values of~$\Lambda$, defined in
  Eqn.~\eqref{eq:lambdaeqn} for different observers located at~$r
  \simeq 2$, $10$, $20$, $30$, $40$, $50$ and~$100 M$ are less
  than~$1$, showing that the effects of the tail play an important
  role during QNM ringing, especially at intermediate and late times,
  since generally~$\Lambda$ decreases with increasing
  time. \textit{Right}: A comparison of the numerical signal with
  theoretical contributions from the QNM only and the combined
  contribution of the QNMs and tail. The inset plot shows the
  pointwise percentage error in the two cases.}
\label{fig:qnm2}
\end{figure*}

For each time-series, we perform fits with the number of damped
exponentials in Eqn.~\eqref{eq:qnmmodel} varying from~$k=2$ to~$16$
and record the fundamental mode frequency. If the above conditions are
met, we also record the first overtone. The percentage error for the
real and imaginary parts of the extracted frequencies are then
calculated for different observers and for different values of
$k$. The corresponding contributions from the~$n=0,1$ modes are
calculated from the Green's function and compared with the results of
the fit.

The Green's function predicts that the contribution from the overtones
is significant during the beginning of the signal, which is why
fitting two damped exponentials results in the largest percentage
error in the value for the principal QNM frequency. Despite a few
exceptions when the percentage error is small, as a general trend the
percentage error decreases as the number of exponentials in the
fitting model are increased. With only two exponentials in the model,
the best value for the~$n=0$ mode is obtained by an observer close to
the horizon with the percentage error under~$10 \%$. In general, with
the addition of~$6$ or more terms in the model, the error for the
fundamental mode can be kept within~$1 \%$ for both the real and
imaginary parts, irrespective of the observer chosen. We also compared
the~$n=0$ mode generated by the fitting algorithm and the Green's
function and found to be in good agreement, with the error between the
two amplitudes at the beginning to be~$< 10 \%$ for most cases.

\subsubsection{Overtones with generic data}

The investigation of overtone modes with generic initial data is less
successful. We choose the same set of simulations and perform the data
analysis using the same methods as the previous section. It is
challenging to reliably extract the first overtone in all cases
because some of the extracted frequencies fail to satisfy the
consistency test for a pair of damped exponentials mentioned
before. In this case, a model with more damped exponentials will not
necessarily lead to a more accurate estimation of the first overtone,
but more than~$4$ damped exponentials are needed for reliable
extraction. The accuracy of the frequency extracted is not highly
dependent on the position of the observer, although the frequency
extracted is seen to be more accurate for observers close to the
horizon. As a general rule, the imaginary part of the frequency is
extracted more accurately than the real part. Even then, for generic
initial data the percentage errors for both the real and imaginary
part of the frequency are too large for them to be of real use. The
best case we observe for our data set is~$< 3 \%$ error in both the
real and imaginary part of the first overtone. The large error or the
inability to detect the first overtone can be attributed to the
limitations of the fitting algorithm, short duration of ringing and
the presence of a significant contribution from the backscattering
during intermediate and late ringdown.

The shortcomings of the linear fitting method may be improved by using
a non-linear algorithm with an improved model incorporating the tail
while the short ringing time may be improved by using specialized
initial data which enhances the duration of ringing. All of this is
discussed in the rest of the paper. A representative fit for the
fundamental mode and the first overtone is shown the top row, left of
Fig.~\ref{fig:qnm1} while on the top right we show the error in
estimating the~$n=0$ QNM frequencies for various positions of the
observer and various number of terms considered.

\subsubsection{QNMs with specialized data}

After limited success in extracting overtone modes with generic
initial data, we wish to prepare specialized data which allows for
more accurate measurement of the first overtone. A naive observation
here is that the data analysis algorithm and subsequently the
parameter estimation works better with a larger number of ringdown
cycles. An elementary way to achieve this is by evolving a pure
overtone type initial data. The alternative approach, which we
describe here, is to approximate the ringing time from the Green's
function. While it is not possible to infer the parameters of the
initial data by specifying a ringing duration~$\Delta t_\textrm{QNM}$,
the converse is easily achieved from combining the results of the QNM
and tail components of the Green's function.  The basic prescription
is outlined below:
\begin{enumerate}[1.]
\item The first step to estimate the duration of ringing is to choose
  a starting time~$t_\textrm{i}$ for ringing. For an asymptotic
  observer, this is the time taken by the ingoing part of the initial
  data to interact with the peak of the scattering potential near the
  light ring and propagate outwards towards the observer. The starting
  time can be intuitively approximated as
\begin{align}
	t_\textrm{i} &\approx r_0 + 5 \sigma + r + 4 M \log(r - 2 M) \nonumber
        \\
        &\quad - 6 M + 4 M \log M.
\end{align}
For observers close to the horizon, a more simple expression can be
obtained,
\begin{align}
	t_\textrm{i} \approx r_0 + 5 \sigma - 2 M.
\end{align}
Here~$r_0$ is the peak of the Gaussian with a standard
deviation~$\sigma / \sqrt{2}$.
\item An appropriate duration for the search $t_\textrm{f}$ is then
  chosen, assuming that the effects of the tail dominate over the QNM
  ringing before this time. In our searches, we choose~$t_\textrm{f} =
  t_\textrm{i} + 300 M$.
\item The QNM contribution to the signal, upto the first three terms
  in Eqn.~\eqref{eq:qnmsum} and the tail contribution to the signal,
  upto the first 15 terms in Eqn.~\eqref{eq:BCGF2} is evaluated
  over~$\left[ t_\textrm{i} , t_\textrm{f} \right]$, for a specific
  choice of initial data.
\item The modulus of the QNM sum amplitude decays linearly and
  intersects with the tail amplitude at time~$t_\textrm{q}$, which we
  shall consider at the end time of ringing. The approximate ringing
  time is taken to be~$\Delta t_\textrm{QNM} \approx t_\textrm{q} -
  t_\textrm{i}$. Some estimates for the approximate ringing time from
  the Green's function are given in the bottom left
  of~Fig.~\ref{fig:qnm1}.
\end{enumerate}
As a test for this method, we perform a brief comparison between
Gaussian II and sine-Gaussian type initial data. For Gaussian II data,
we estimate~$\Delta t_\textrm{QNM}$ for~$5$ different values
of~$\sigma = 1, 3, 5, 7, 9$. The same is used for sine-Gaussian data
with~$\omega = 0.1, 0.5, 1, 3, 5$ for each~$\sigma$. In both cases,
the Gaussian is centered around~$r_0 = 100 M$ and the observer is
positioned at $r \approx 100 M$. We observe an appreciable variation
in~$\Delta t_\textrm{QNM}$ when sine-Gaussians are used, in fact with
suitable choice of parameters, $\Delta t_\textrm{QNM} \sim 250 M$
which is about~$5$ times what we can achieve with Gaussian II data. We
must note here that although such long duration ringing may be seen by
observers far away from the event horizon in principle, it is hardly
the case in practice owing to the constraints from numerical
noise. This technical problem could be redressed by assigning more
memory for floating point numbers in \verb|bamps|. To illustrate the
point that the QNM frequencies can be extracted from the data more
reliably, we consider two simulations, one with Gaussian II type data
with parameters~$A = 10^4 M$, $\sigma = 7 M$, $r_0 = 100 M$ and
another with sine-Gaussian data with parameters~$A = 10^4 M $, $\sigma
= 5 M$, $r_0 = 100 M$, $\omega M = 1$. An observer is placed at~$r
\approx 100 M$ and a fit of damped sinusoids is performed on the QNM
part of the time-series extracted in both cases. A plot of the
signals, as seen in the bottom right of Fig.~\ref{fig:qnm1}, shows a
very short ringdown phase in the first signal, labeled as \verb|B:7|
while a much longer ringdown phase is observed in the second signal,
labeled as \verb|C:5|. A longer ringdown signal enables QNM
frequencies to be estimated more accurately far away from the black
hole with some estimates given in table~\ref{tab:overtones}.

\begin{table}[t!]
	\centering
	\begin{tabular}{cccc}
		Simulation& $n$ & $\omega_{n}$ & $\%$ error \\ 
		\hline \hline
		B:7 & 0 & $0.1179-0.1039i$ & $(6.71, 0.99)$ \\
		  & 1 & NA & NA \\
		C:5 & 0 & 0.1105-0.1050i & (0.05, 0.06) \\
		  & 1 & 0.0912-0.3630i & (5.96, 4.31) \\
		\hline
	\end{tabular}
	\caption{Estimated values of the~$n = 0$ and~$n = 1$ for
          generic and special initial data as measured by an observer
          at~$r \simeq 100 M$.}\label{tab:overtones}
\end{table}

In our case, there is an improvement of two orders of magnitude for
the~$n=0$ mode, which is impressive given that all~$l=0$ modes are
damped away fairly quickly.

\subsubsection{Importance of tails during ringdown}

\begin{figure*}[t]
\centering
\includegraphics[width=\textwidth]{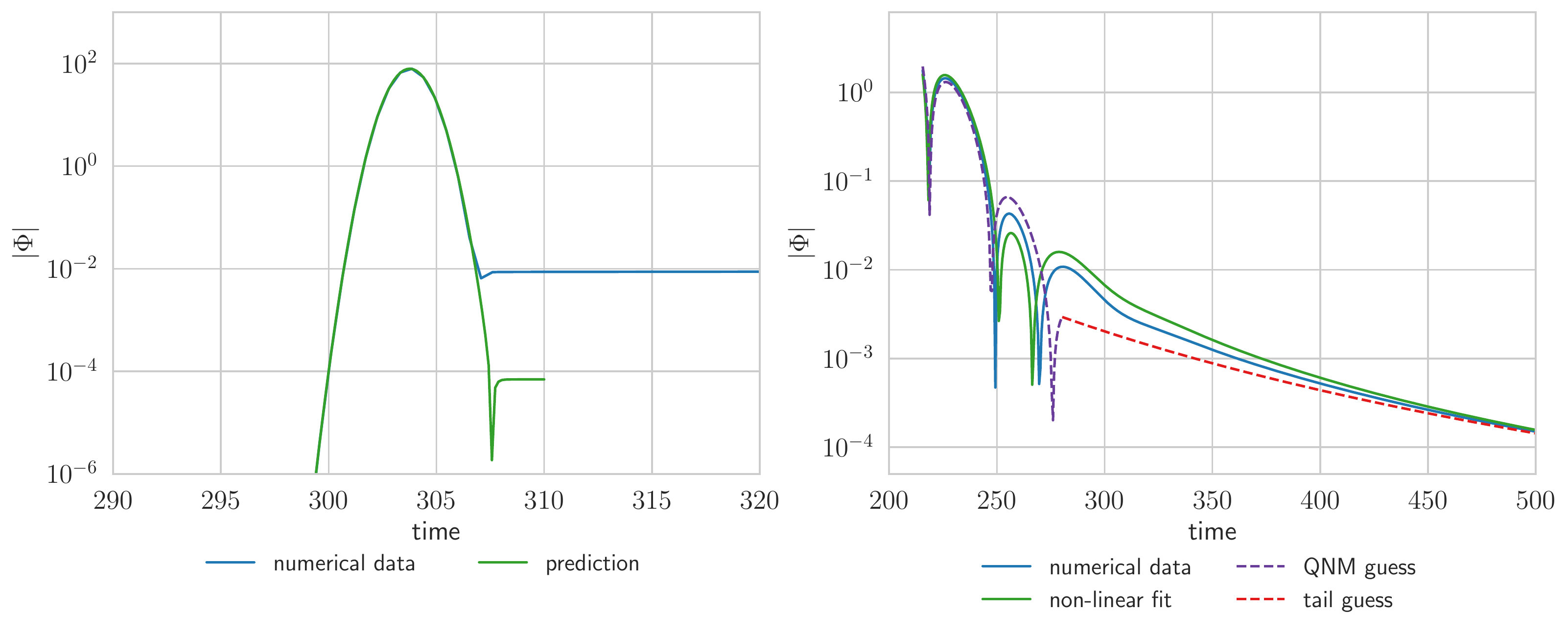}
\caption{\textit{Left}: A fit for the scalar field time series as seen
  by an observer near the horizon ($r \approx 2 M$) with the model in
  Eqn.~\eqref{eq:lmfit} using the Levenberg-Marquardt algorithm. The
  separate QNM and tail guesses are also shown. \textit{Right}: A
  comparison between the contribution of the high frequency arc to the
  signal from the approximate Green's function and the numerical data
  for Gaussian~I type initial data with the observer at~$r \simeq 500
  M$.}\label{fig:directplot}
\end{figure*}

We now investigate the effect of the branch cut to the signal during
QNM ringing. To do this, we compute the overtone and the approximate
tail contribution for the entire duration of the `ringing signal'. The
tail contribution is approximated by extending the low frequency
expressions in Eqn.~\eqref{eq:BCGF2}, evaluated up to the first~$15$
terms to earlier times and the QNM contribution is computed from the
sum of the contribution of the first three modes. We then calculate
the difference between the numerical data~$\Phi_\textrm{b}$ and the
mode sum~$\Phi_\textrm{q}$ and between the numerical data minus the
approximate tail contribution~$\Phi_\textrm{t}$ and the mode sum. The
modulus of the ratio of these two quantities
\begin{align} \label{eq:lambdaeqn}
  \Lambda = \left| \frac{\Phi_\textrm{b} - \Phi_\textrm{t} - \Phi_\textrm{q}}
          {\Phi_\textrm{b} - \Phi_\textrm{q}} \right|,
\end{align}
is observed as a function of time. For all simulations
considered,~$\Lambda$ is seen to be less than~$1$ and in general
decreases with increasing time. This can be seen in the left of
Fig.~\ref{fig:qnm2}. This demonstrates that the contribution from the
tail becomes important during intermediate and late time ringing and
should be considered in the fitting model along with the damped
sinusoids for better extraction of the QNM frequencies. As a proof of
concept, we fit damped sinusoids to~$\Phi_\textrm{b} -
\Phi_\textrm{t}$ using the linear fitting strategy described before
to~$168$ signals and observe an improvement in the percentage error
for the principal QNM frequency in~$\sim 69 \%$ cases while the
improvement in measuring the first overtone is seen in~$~ 33 \%$ cases
for generic initial data. On the right of Fig.~\ref{fig:qnm2} we show
a representative plot where both contributions of the overtones and
the tail are considered during ringing.

\subsection{Approximation of pre-ringdown} \label{subsection:preringdown}

We make a comparison between the leading order contribution from the
high frequency arc and the numerical data and find good agreement with
the numerics at early times but the flat space approximation rapidly
fails at later times. This shortcoming can be redressed by considering
higher order terms in the high frequency approximation of the
Whittaker functions. In the left of Fig.~\ref{fig:directplot} we
display a comparison plot for Gaussian I type data.

\section{Discussion and conclusions}\label{Section:Conclusions}

Motivated both by gravitational wave astronomy and by pure theory, the
principal objective of this paper was to help facilitate, in the near
future, a comparison between linear and non-linear perturbation theory
by extending the Green's function approach to arbitrary
horizon-penetrating coordinates. This allows us to find the dynamic
excitation amplitude of each QNM excited for any observer outside the
event horizon. This was achieved by generalizing the computations
of~\cite{DV} for QNMs in Eddington-Finkelstein coordinates to
arbitrary horizon penetrating coordinates, and computing the exact
Green's function from solutions of the CHE. Under the approximation
that the observer is far away from the event horizon, the solutions of
the asymptotic form of CHE are just solutions of a Whittaker
equation. The resulting expressions for the asymptotic Green's
function are much easier to handle. They were then used to compute the
dominant contribution from the high frequency arc as well as the
contributions from the branch cut at low, intermediate and high
frequencies. The late time contribution from the branch cut gives rise
to Price's tail law. These results were then put to the test using the
new \verb|scalarfield| project inside \verb|bamps|, in which a single
Schwarzschild black hole is perturbed by different configurations of a
spherically symmetric massless scalar field.

Besides a verification of our mathematical results, the numerical
experiments also show that the first overtone mode may not be reliably
extracted from generic initial data for observers far away from the
black hole. However, by using specialized initial data we were able to
increase the duration of ringing almost threefold, thereby extracting
the frequencies of the fundamental mode and the first overtone more
accurately. We also found that the branch cut contributes
significantly during intermediate and late ringing, and must be taken
into account in the data analysis model. It is therefore sensible to
consider a data analysis model for QNM ringing which also incorporates
the effect of the branch cut, as given for example by,
\begin{align} \label{eq:lmfit}
\Phi_{m} = \sum_{j=0}^{2} A_j e^{-\alpha_j t}\sin (\omega_j t + \phi_j)
+ \sum_{k=0}^{4} \frac{B_k}{t^{3+k}}.
\end{align}
In our experiments with this model we found that at least~$3$ damped
sinusoids and~$5$ tail terms are needed for an accurate representation
of the signal. Additionally, the starting time for the signal has to
be determined by an additional parameter.

The Levenberg-Marquardt non-linear least squares technique may be used
to fit the model to the data. We find, however, that the method may
fail to converge if initial guesses for the parameters are far away
from their correct values. Our strategy to obtain good parameters for
the tail terms is to isolate the tail signal and perform the tail
analysis separately while for the QNM parameters, we obtain good
initial values with the matrix pencil method. All of these parameters
are then used as initial guesses while fitting for the entire signal
for different starting times of the fit. In the right of
Fig.~\ref{fig:directplot}, a demonstration of such a fit is given.

Several improvements are possible on the present approach. While we
see that the asymptotic expressions for the tail work well, even for
observers close to the event horizon, an exact Green's function for
the branch cut may also be obtained using the solutions of the CHE
built along the lines of the MST approach~\cite{MSTa, MSTb}. Our
approximation for the contribution of the high frequency arc fails to
account for the subdominant terms for which a more nuanced approach
for handling high frequency approximations of the Whittaker function
is necessary. It is known that solutions to the Teukolsky master
equation can be written down in terms of the confluent Heun
equation~\cite{Fiziev_2010}, so another natural extension to this work
would be to consider the spin-$1$ and spin-$2$ cases. Our comparison
between the linear results and full non-linear theory is ongoing and
will be presented separately.

\acknowledgments

We are grateful to Nils Andersson, Emanuele Berti, Sukanta Bose,
Plamen Fiziev, Edgar Gasperin, Shalabh Gautam, Praveer Krishna
Gollapudi, Rodrigo Panosso Macedo, Volker Perlick, Dennis Philipp,
Andrzej Rostworowski and Chiranjeeb Singha for helpful discussions and
feedback on the manuscript. We are particularly indebted to Sanjeev
Dhurandhar for his continuous support and encouragement, and without
whom this project would not have been possible. MKB and KRN
acknowledges support from the Ministry of Human Resource Development
(MHRD), India, IISER Kolkata and the Center of Excellence in Space
Sciences (CESSI), India, the Newton-Bhaba partnership between LIGO
India and the University of Southampton, the Navajbai Ratan Tata Trust
grant and the Visitors' Programme at the Inter-University Centre for
Astronomy and Astrophysics (IUCAA), Pune. CESSI, a multi-institutional
Center of Excellence established at IISER Kolkata is funded by the
MHRD under the Frontier Areas of Science and Technology (FAST) scheme.
DH gratefully acknowledges support offered by IUCAA, Pune, where part
of this work was completed. The work was partially supported by the
FCT (Portugal) IF Program~IF/00577/2015, Project~No.~UIDB/00099/2020
and PTDC/MAT-APL/30043/2017.

\bibliography{horizon_qnm_tail.bbl}{}

\begin{thebibliography}{10}

\bibitem{RW}
Tullio Regge and John~A. Wheeler.
\newblock Stability of a {S}chwarzschild singularity.
\newblock {\em Phys. Rev.}, 108:1063--1069, Nov 1957.

\bibitem{Zerilli}
Frank~J. Zerilli.
\newblock Gravitational field of a particle falling in a {S}chwarzschild
  geometry analyzed in tensor harmonics.
\newblock {\em Phys. Rev. D}, 2:2141--2160, Nov 1970.

\bibitem{Vishu}
C.~V. {Vishveshwara}.
\newblock {Scattering of gravitational radiation by a {S}chwarzschild
  black-hole}.
\newblock {\em \nat}, 227:936--938, August 1970.

\bibitem{BertiReview}
Emanuele Berti, Vitor Cardoso, and Andrei~O Starinets.
\newblock Quasinormal modes of black holes and black branes.
\newblock {\em Classical and Quantum Gravity}, 26(16):163001, 2009.

\bibitem{Nollert}
Hans-Peter Nollert.
\newblock Quasinormal modes: {T}he characteristic `sound' of black holes and
  neutron stars.
\newblock {\em Classical and Quantum Gravity}, 16(12):R159, 1999.

\bibitem{Kokkotas1999}
Kostas~D. Kokkotas and Bernd~G. Schmidt.
\newblock Quasi-normal modes of stars and black holes.
\newblock {\em Living Reviews in Relativity}, 2(1):2, Sep 1999.

\bibitem{Konoplya}
R.~A. Konoplya and Alexander Zhidenko.
\newblock Quasinormal modes of black holes: {F}rom astrophysics to string
  theory.
\newblock {\em Rev. Mod. Phys.}, 83:793--836, Jul 2011.

\bibitem{LIGO1}
B.~P. Abbott et~al.
\newblock Observation of gravitational waves from a binary black hole merger.
\newblock {\em Phys. Rev. Lett.}, 116:061102, Feb 2016.

\bibitem{LIGO2}
B.~P. Abbott et~al.
\newblock G{W}151226: Observation of gravitational waves from a 22-solar-mass
  binary black hole coalescence.
\newblock {\em Phys. Rev. Lett.}, 116:241103, Jun 2016.

\bibitem{LIGO3}
B.~P. Abbott et~al.
\newblock G{W}170104: Observation of a 50-solar-mass binary black hole
  coalescence at redshift 0.2.
\newblock {\em Phys. Rev. Lett.}, 118:221101, Jun 2017.

\bibitem{LIGO4}
B.~P. Abbott et~al.
\newblock G{W}170608: Observation of a 19 solar-mass binary black hole
  coalescence.
\newblock {\em The Astrophysical Journal Letters}, 851(2):L35, 2017.

\bibitem{LIGO5}
B.~P. Abbott et~al.
\newblock G{W}170814: A three-detector observation of gravitational waves from
  a binary black hole coalescence.
\newblock {\em Phys. Rev. Lett.}, 119:141101, Oct 2017.

\bibitem{LIGO6}
B.~P. Abbott et~al.
\newblock Multi-messenger observations of a binary neutron star merger.
\newblock {\em The Astrophysical Journal Letters}, 848(2):L12, 2017.

\bibitem{LIGO7}
B.~P. Abbott et~al.
\newblock Tests of general relativity with {G}{W}150914.
\newblock {\em Phys. Rev. Lett.}, 116:221101, May 2016.

\bibitem{LIGO8}
B.~P. Abbott et~al.
\newblock {Tests of general relativity with {G}{W}170817}.
\newblock {\em arXiv e-prints}, November 2018.

\bibitem{Li1}
T.~G.~F. Li et~al.
\newblock Towards a generic test of the strong field dynamics of general
  relativity using compact binary coalescence.
\newblock {\em Phys. Rev. D}, 85:082003, Apr 2012.

\bibitem{Li2}
M.~Agathos et~al.
\newblock Tiger: A data analysis pipeline for testing the strong-field dynamics
  of general relativity with gravitational wave signals from coalescing compact
  binaries.
\newblock {\em Phys. Rev. D}, 89:082001, Apr 2014.

\bibitem{Li3}
Jeroen Meidam et~al.
\newblock Parametrized tests of the strong-field dynamics of general relativity
  using gravitational wave signals from coalescing binary black holes: {F}ast
  likelihood calculations and sensitivity of the method.
\newblock {\em Phys. Rev. D}, 97:044033, Feb 2018.

\bibitem{Israel}
Werner Israel.
\newblock Event horizons in static vacuum space-times.
\newblock {\em Phys. Rev.}, 164:1776--1779, Dec 1967.

\bibitem{Israel2}
B.~Carter.
\newblock Axisymmetric black hole has only two degrees of freedom.
\newblock {\em Phys. Rev. Lett.}, 26:331--333, Feb 1971.

\bibitem{Isi}
Maximiliano Isi, Matthew Giesler, Will~M. Farr, Mark~A. Scheel, and Saul~A.
  Teukolsky.
\newblock Testing the no-hair theorem with {G}{W}150914.
\newblock {\em Phys. Rev. Lett.}, 123:111102, Sep 2019.

\bibitem{Pullin1}
Reinaldo~J. Gleiser, Carlos~O. Nicasio, Richard~H. Price, and Jorge Pullin.
\newblock Colliding black holes: How far can the close approximation go?
\newblock {\em Phys. Rev. Lett.}, 77:4483--4486, Nov 1996.

\bibitem{Pullin2}
Reinaldo~J Gleiser, Carlos~O Nicasio, Richard~H Price, and Jorge Pullin.
\newblock Second-order perturbations of a {S}chwarzschild black hole.
\newblock {\em Classical and Quantum Gravity}, 13(10):L117, 1996.

\bibitem{Pullin3}
Carlos~O. Nicasio, Reinaldo~J. Gleiser, Richard~H. Price, and Jorge Pullin.
\newblock Collision of boosted black holes: {S}econd order close limit
  calculations.
\newblock {\em Phys. Rev. D}, 59:044024, Jan 1999.

\bibitem{Pullin4}
Reinaldo~J. Gleiser, Carlos~O. Nicasio, Richard~H. Price, and Jorge Pullin.
\newblock Evolving the {B}owen-{Y}ork initial data for spinning black holes.
\newblock {\em Phys. Rev. D}, 57:3401--3407, Mar 1998.

\bibitem{Pullin5}
Reinaldo~J. Gleiser, Carlos~O. Nicasio, Richard~H. Price, and Jorge Pullin.
\newblock Gravitational radiation from {S}chwarzschild black holes: the
  second-order perturbation formalism.
\newblock {\em Physics Reports}, 325(2):41 -- 81, 2000.

\bibitem{nonlinear1}
Hirotada Okawa, Helvi Witek, and Vitor Cardoso.
\newblock Black holes and fundamental fields in numerical relativity: {I}nitial
  data construction and evolution of bound states.
\newblock {\em Phys. Rev. D}, 89:104032, May 2014.

\bibitem{nonlinear2}
Nicolas Sanchis-Gual, Juan~Carlos Degollado, Pedro~J. Montero, and Jos{\'e}~A.
  Font.
\newblock Quasistationary solutions of self-gravitating scalar fields around
  black holes.
\newblock {\em Phys. Rev. D}, 91:043005, Feb 2015.

\bibitem{nonlinear3}
Nicolas Sanchis-Gual, Juan~Carlos Degollado, Paula Izquierdo, Jos{\'e}~A. Font,
  and Pedro~J. Montero.
\newblock Quasistationary solutions of scalar fields around accreting black
  holes.
\newblock {\em Phys. Rev. D}, 94:043004, Aug 2016.

\bibitem{nonlinear4}
Robert Benkel, Thomas~P. Sotiriou, and Helvi Witek.
\newblock Dynamical scalar hair formation around a {S}chwarzschild black hole.
\newblock {\em Phys. Rev. D}, 94:121503, Dec 2016.

\bibitem{Hiroyuki}
Hiroyuki Nakano and Kunihito Ioka.
\newblock Second-order quasinormal mode of the {S}chwarzschild black hole.
\newblock {\em Phys. Rev. D}, 76:084007, Oct 2007.

\bibitem{Bizon2009}
Piotr Bizo{\'{n}}, Tadeusz Chmaj, and Andrzej Rostworowski.
\newblock Late-time tails of a self-gravitating massless scalar field,
  revisited.
\newblock {\em Classical and Quantum Gravity}, 26(17):175006, Aug 2009.

\bibitem{bamps1}
Bernd Br{\"u}gmann.
\newblock A pseudospectral matrix method for time-dependent tensor fields on a
  spherical shell.
\newblock {\em Journal of Computational Physics}, 235:216 -- 240, 2013.

\bibitem{bamps2}
David Hilditch, Andreas Weyhausen, and Bernd Br{\"u}gmann.
\newblock Pseudospectral method for gravitational wave collapse.
\newblock {\em Phys. Rev. D}, 93:063006, Mar 2016.

\bibitem{bamps3}
Marcus Bugner, Tim Dietrich, Sebastiano Bernuzzi, Andreas Weyhausen, and Bernd
  Br{\"u}gmann.
\newblock Solving 3{D} relativistic hydrodynamical problems with weighted
  essentially nonoscillatory discontinuous {G}alerkin methods.
\newblock {\em Phys. Rev. D}, 94:084004, Oct 2016.

\bibitem{bamps4}
David Hilditch, Enno Harms, Marcus Bugner, Hannes R{\"u}ter, and Bernd
  Br{\"u}gmann.
\newblock The evolution of hyperboloidal data with the dual foliation
  formalism: mathematical analysis and wave equation tests.
\newblock {\em Classical and Quantum Gravity}, 35(5):055003, 2018.

\bibitem{bamps5}
David Hilditch, Andreas Weyhausen, and Bernd Br{\"u}gmann.
\newblock Evolutions of centered {B}rill waves with a pseudospectral method.
\newblock {\em Phys. Rev. D}, 96:104051, Nov 2017.

\bibitem{bamps6}
Hannes~R. R{\"u}ter, David Hilditch, Marcus Bugner, and Bernd Br{\"u}gmann.
\newblock Hyperbolic relaxation method for elliptic equations.
\newblock {\em Phys. Rev. D}, 98:084044, Oct 2018.

\bibitem{bamps7}
Andreas Schoepe, David Hilditch, and Marcus Bugner.
\newblock Revisiting hyperbolicity of relativistic fluids.
\newblock {\em Phys. Rev. D}, 97:123009, Jun 2018.

\bibitem{ansorgrodrigo}
Marcus Ansorg and Rodrigo~Panosso Macedo.
\newblock Spectral decomposition of black-hole perturbations on hyperboloidal
  slices.
\newblock {\em Phys. Rev. D}, 93:124016, Jun 2016.

\bibitem{Campanelli_2001}
Manuela Campanelli, Gaurav Khanna, Pablo Laguna, Jorge Pullin, and Michael~P
  Ryan.
\newblock Perturbations of the {K}err spacetime in horizon-penetrating
  coordinates.
\newblock {\em Classical and Quantum Gravity}, 18(8):1543--1554, March 2001.

\bibitem{PhysRevD.64.084016}
Olivier Sarbach and Manuel Tiglio.
\newblock Gauge-invariant perturbations of {S}chwarzschild black holes in
  horizon-penetrating coordinates.
\newblock {\em Phys. Rev. D}, 64:084016, Sep 2001.

\bibitem{LeaverPRD}
Edward~W. Leaver.
\newblock Spectral decomposition of the perturbation response of the
  {S}chwarzschild geometry.
\newblock {\em Phys. Rev. D}, 34:384--408, Jul 1986.

\bibitem{Andersson1995}
N.~{Andersson}.
\newblock {Excitation of {S}chwarzschild black-hole quasinormal modes}.
\newblock {\em \prd}, 51:353--363, January 1995.

\bibitem{Andersson1997}
Nils Andersson.
\newblock Evolving test fields in a black-hole geometry.
\newblock {\em Phys. Rev. D}, 55:468--479, Jan 1997.

\bibitem{Berti2006}
Emanuele Berti and Vitor Cardoso.
\newblock Quasinormal ringing of {K}err black holes: The excitation factors.
\newblock {\em Phys. Rev. D}, 74:104020, Nov 2006.

\bibitem{CaltechGF}
Huan Yang, Fan Zhang, Aaron Zimmerman, and Yanbei Chen.
\newblock Scalar {G}reen function of the {K}err spacetime.
\newblock {\em Phys. Rev. D}, 89:064014, Mar 2014.

\bibitem{Zhang}
Zhongyang Zhang, Emanuele Berti, and Vitor Cardoso.
\newblock Quasinormal ringing of {K}err black holes. {I}{I}. {E}xcitation by
  particles falling radially with arbitrary energy.
\newblock {\em Phys. Rev. D}, 88:044018, Aug 2013.

\bibitem{PhysRevD.84.104002}
Sam~R. Dolan and Adrian~C. Ottewill.
\newblock Wave propagation and quasinormal mode excitation on {S}chwarzschild
  spacetime.
\newblock {\em Phys. Rev. D}, 84:104002, Nov 2011.

\bibitem{PhysRevD.38.1040}
Yonghe Sun and Richard~H. Price.
\newblock Excitation of quasinormal ringing of a {S}chwarzschild black hole.
\newblock {\em Phys. Rev. D}, 38:1040--1052, Aug 1988.

\bibitem{Frolov:1998wf}
V.P. Frolov and I.D. Novikov, editors.
\newblock {\em {Black hole physics: {B}asic concepts and new developments}},
  volume~96.
\newblock 1998.

\bibitem{Fiziev1}
Plamen~P Fiziev.
\newblock Exact solutions of {R}egge-{W}heeler equation and quasi-normal modes
  of compact objects.
\newblock {\em Classical and Quantum Gravity}, 23(7):2447, 2006.

\bibitem{Fiziev2}
Plamen~P Fiziev.
\newblock Classes of exact solutions to the {T}eukolsky master equation.
\newblock {\em Classical and Quantum Gravity}, 27(13):135001, 2010.

\bibitem{Fiziev3}
Plamen~P Fiziev.
\newblock Novel relations and new properties of confluent {H}eun's functions
  and their derivatives of arbitrary order.
\newblock {\em Journal of Physics A: Mathematical and Theoretical},
  43(3):035203, 2010.

\bibitem{London}
Lionel London, Deirdre Shoemaker, and James Healy.
\newblock Modeling ringdown: {B}eyond the fundamental quasinormal modes.
\newblock {\em Phys. Rev. D}, 90:124032, Dec 2014.

\bibitem{Giesler}
Matthew Giesler, Maximiliano Isi, Mark~A. Scheel, and Saul~A. Teukolsky.
\newblock Black hole ringdown: {T}he importance of overtones.
\newblock {\em Phys. Rev. X}, 9:041060, Dec 2019.

\bibitem{hortacsu}
M.~Hortacsu.
\newblock {\em Mathematical Physics: {H}eun functions and their uses in
  physics}, pages 23--39.

\bibitem{Fiziev4}
Plamen Fiziev and Denitsa Staicova.
\newblock Application of the confluent {H}eun functions for finding the
  quasinormal modes of nonrotating black holes.
\newblock {\em Phys. Rev. D}, 84:127502, Dec 2011.

\bibitem{DV}
Dennis {Philipp} and Volker {Perlick}.
\newblock {Schwarzschild radial perturbations in {E}ddington-{F}inkelstein and
  {P}ainlev{\'e}-{G}ullstrand coordinates}.
\newblock {\em International Journal of Modern Physics D}, 24:1542006, May
  2015.

\bibitem{VIEIRA201414}
H.S. Vieira, V.B. Bezerra, and C.R. Muniz.
\newblock Exact solutions of the {K}lein-{G}ordon equation in the
  {K}err-{N}ewman background and {H}awking radiation.
\newblock {\em Annals of Physics}, 350:14 -- 28, 2014.

\bibitem{FizievInterior1}
Plamen~P. {Fiziev}.
\newblock {On the Exact Solutions of the {R}egge-{W}heeler Equation in the
  {S}chwarzschild black hole interior}.
\newblock {\em arXiv e-prints}, pages gr--qc/0603003, Mar 2006.

\bibitem{ishkhanyan}
T.~A. Ishkhanyan and A.~M. Ishkhanyan.
\newblock Expansions of the solutions to the confluent {H}eun equation in terms
  of the {K}ummer confluent {H}ypergeometric functions.
\newblock {\em AIP Advances}, 4(8):087132, 2014.

\bibitem{NIST:DLMF}
{\it NIST Digital Library of Mathematical Functions}.
\newblock http://dlmf.nist.gov/, Release 1.0.20 of 2018-09-15.
\newblock F.~W.~J. Olver, A.~B. {Olde Daalhuis}, D.~W. Lozier, B.~I. Schneider,
  R.~F. Boisvert, C.~W. Clark, B.~R. Miller and B.~V. Saunders, eds.

\bibitem{SlavyanovBook}
Sergei~Yu. Slavyanov and Wolfgang Lay.
\newblock {\em Special functions}.
\newblock Oxford Mathematical Monographs. Oxford University Press, Oxford,
  2000.
\newblock A unified theory based on singularities, With a foreword by Alfred
  Seeger, Oxford Science Publications.

\bibitem{Olver}
F.W.J. Olver.
\newblock Differential equations with irregular singularities; {B}essel and
  confluent {H}ypergeometric functions.
\newblock In F.W.J. Olver, editor, {\em Asymptotics and Special Functions},
  pages 229 -- 278. Academic Press, 1974.

\bibitem{MSTa}
Shuhei Mano, Hisao Suzuki, and Eiichi Takasugi.
\newblock Analytic solutions of the {T}eukolsky equation and their low
  frequency expansions.
\newblock {\em Progress of Theoretical Physics}, 95(6):1079--1096, 1996.

\bibitem{MSTb}
Shuhei Mano, Hisao Suzuki, and Eiichi Takasugi.
\newblock Analytic solutions of the {R}egge-{W}heeler equation and the
  post-{M}inkowskian expansion.
\newblock {\em Progress of Theoretical Physics}, 96(3):549--565, 1996.

\bibitem{Casals1}
Marc Casals and Adrian Ottewill.
\newblock High-order tail in {S}chwarzschild spacetime.
\newblock {\em Phys. Rev. D}, 92:124055, Dec 2015.

\bibitem{Leaver2}
E.~W. {Leaver}.
\newblock {Solutions to a generalized spheroidal wave equation: {T}eukolsky's
  equations in general relativity, and the two-center problem in molecular
  quantum mechanics}.
\newblock {\em Journal of Mathematical Physics}, 27:1238--1265, May 1986.

\bibitem{Leaver1}
E.~W. Leaver.
\newblock An analytic representation for the quasi-normal modes of {K}err black
  holes.
\newblock {\em Proceedings of the Royal Society of London A: Mathematical,
  Physical and Engineering Sciences}, 402(1823):285--298, 1985.

\bibitem{FizievQNMsPRD}
Plamen Fiziev and Denitsa Staicova.
\newblock Application of the confluent {H}eun functions for finding the
  quasinormal modes of nonrotating black holes.
\newblock {\em Phys. Rev. D}, 84:127502, Dec 2011.

\bibitem{szpak}
Nikodem {Szpak}.
\newblock {Quasinormal mode expansion and the exact solution of the Cauchy
  problem for wave equations}.
\newblock {\em arXiv e-prints}, pages gr--qc/0411050, Nov 2004.

\bibitem{GradshteynRyzhik}
I.~S. Gradshteyn and I.~M. Ryzhik.
\newblock {\em Table of integrals, series, and products}.
\newblock Elsevier/Academic Press, Amsterdam, eighth edition, 2015.

\bibitem{PriceTail}
Richard~H. Price.
\newblock Nonspherical perturbations of relativistic gravitational collapse.
  {I}. {S}calar and gravitational perturbations.
\newblock {\em Phys. Rev. D}, 5:2419--2438, May 1972.

\bibitem{alcubierre}
Miguel Alcubierre.
\newblock Initial data.
\newblock In {\em Introduction to 3+1 Numerical Relativity}. Oxford University
  Press, 2012.

\bibitem{BS}
Thomas~W. Baumgarte and Stuart~L. Shapiro.
\newblock In {\em Numerical Relativity: {S}olving Einstein's Equations on the
  Computer}. Cambridge University Press, 2010.

\bibitem{cartoon1}
M.~Alcubierre, S.~R. Brandt, B.~Br{\"u}gmann, D.~Holz, E.~Seidel, R.~Takahashi,
  and J.~Thornburg.
\newblock Symmetry without symmetry: Numerical simulation of axisymmetric
  systems using cartesian grids.
\newblock {\em Int. J. Mod. Phys. D}, 10(3):273--289, 2001.

\bibitem{cartoon2}
Frans Pretorius.
\newblock Numerical relativity using a generalized harmonic decomposition.
\newblock {\em Class. Quant. Grav.}, 22:425--451, 2005.

\bibitem{Harms_2013}
Enno Harms, Sebastiano Bernuzzi, and Bernd Br{\"u}gmann.
\newblock Numerical solution of the 2+1 {T}eukolsky equation on a hyperboloidal
  and horizon penetrating foliation of {K}err and application to late-time
  decays.
\newblock {\em Classical and Quantum Gravity}, 30(11):115013, may 2013.

\bibitem{lmfit}
Matt Newville et~al.
\newblock lmfit/lmfit-py 0.9.13, April 2019.

\bibitem{Hua}
Y.~Hua and T.~K. Sarkar.
\newblock Matrix pencil method for estimating parameters of exponentially
  damped/undamped sinusoids in noise.
\newblock {\em IEEE Transactions on Acoustics, Speech, and Signal Processing},
  38(5):814--824, May 1990.

\bibitem{BertiDA}
Emanuele Berti, Vitor Cardoso, Jos\'e~A. Gonz\'alez, and Ulrich Sperhake.
\newblock Mining information from binary black hole mergers: A comparison of
  estimation methods for complex exponentials in noise.
\newblock {\em Phys. Rev. D}, 75:124017, Jun 2007.

\bibitem{Fiziev_2010}
Plamen~P Fiziev.
\newblock Classes of exact solutions to the {T}eukolsky master equation.
\newblock {\em Classical and Quantum Gravity}, 27(13):135001, may 2010.

\end{thebibliography}

\end{document}